\documentclass[]{spie}  %>>> use for US letter paper
\usepackage{amsmath,amsfonts,amssymb}
\usepackage{graphicx}
\usepackage[colorlinks=true, allcolors=blue]{hyperref}
\usepackage{makecell}
\usepackage{colortbl}
\usepackage{array}
\usepackage{afterpage}

\usepackage[table]{xcolor}

\def\arcdeg{\mbox{$^\circ$}}%
\def\arcsec{\mbox{$^{\prime\prime}$}}%

\definecolor{eric}{rgb}{0.7, 0.7, 1.0}

\title{Curved detectors for future X-ray astrophysics missions}

\author[a]{Eric D.\ Miller}
\author[b]{James A.\ Gregory}
\author[a]{Marshall W.\ Bautz}
\author[b]{Harry R.\ Clark}
\author[b]{Michael Cooper}
\author[b]{Kevan Donlon}
\author[a]{Richard F.\ Foster}
\author[a]{Catherine E.\ Grant}
\author[b]{Mallory Jensen}
\author[a]{Beverly LaMarr}
\author[b]{Renee Lambert}
\author[b]{Christopher Leitz}
\author[a]{Andrew Malonis}
\author[b]{Mo Neak}
\author[a]{Gregory Prigozhin}
\author[b]{Kevin Ryu}
\author[a]{Benjamin Schneider}
\author[b]{Keith Warner}
\author[b]{Douglas J.\ Young}
\author[c]{William W.\ Zhang}

\affil[a]{Kavli Institute for Astrophysics and Space Research, Massachusetts Institute of Technology, Cambridge, MA, USA}
\affil[b]{Lincoln Laboratory, Massachusetts Institute of Technology, Lexington, MA, USA}
\affil[c]{NASA Goddard Space Flight Center, Greenbelt, MD, USA}

\authorinfo{Further author information:\\E.~D.~Miller: E-mail: milleric@mit.edu}

\begin{document} 
\maketitle

\begin{abstract}
Future X-ray astrophysics missions will survey large areas of the sky with unparalleled sensitivity, enabled by lightweight, high-resolution optics. These optics inherently produce curved focal surfaces with radii as small as 2~m, requiring a large area detector system that closely conforms to the curved focal surface. We have embarked on a project using a curved charge-coupled device (CCD) detector technology developed at MIT Lincoln Laboratory to provide large-format, curved detectors for such missions, improving performance and simplifying design. We present the current status of this work, which aims to curve back-illuminated, large-format (5 cm x 4 cm) CCDs to 2.5-m radius and confirm X-ray performance. We detail the design of fixtures and the curving process, and present intial results on curving bare silicon samples and monitor devices and characterizing the surface geometric accuracy. The tests meet our accuracy requirement of $<$5 $\mu$m RMS surface non-conformance for samples of similar thickness to the functional detectors. We finally show X-ray performance measurements of planar CCDs that will serve as a baseline to evaluate the curved detectors. The detectors exhibit low noise, good charge-transfer efficiency, and excellent, uniform spectroscopic performance, including in the important soft X-ray band.
\end{abstract}


\section{INTRODUCTION}
\label{sect:intro}

\subsection{Astrophysics enabled by wide-field, high-spatial-resolution X-ray imaging}

The frontier of high-energy astrophysics in the next decade requires instruments that probe deeply over a wide field of the sky.  Deep X-ray surveys will reveal the seeds of the supermassive black holes that reside at the heart of every galaxy, shaping the evolution of their hosts in ways that are not well-understood, beginning less than a billion years after the Big Bang\cite{Rees1984,Volonteri2010,Inayoshi2020}. Only with sufficient sky coverage will we assemble a statistically useful sample of these seeds and be able to explore how they grow and evolve, along with their host galaxies, into the systems we see today. These studies will capitalize on facilities in other wavebands now coming on-line, and many such future facilities (e.g., the Rubin Observatory, SKA, and LISA) will also identify scores of transient sources that can only be understood with multi-wavelength and multi-messenger observations\cite{Rees1988,Troja2018}. Since the triggers for transients may not be well-localized, especially from gravitational wave observatories, X-ray observatories of the future need to have sufficient grasp to efficiently search a large region of the sky, localize the transient, resolve it from confusing sources, and characterize its emission.

Closer to home, we still have not found the bulk of the baryons in the local Universe, although simulations suggest they should be in a tenuous, hot medium surrounding galaxies and galaxy clusters\cite{Smith2011,Bregman2018}. This material would produce very faint extended X-ray emission, and its challenging sensitivity requirements demand a census of faint, contaminating point sources across a wide field.  This material can also be detected in absorption against bright point sources using dispersive spectroscopy\cite{Smith2020_Arcus,Zhangetal2019}. The absorption lines from highly ionized oxygen and other elements will be very faint, so high sensitivity and high spectral resolution are vital.

These science goals require X-ray instruments with high spatial resolution over a large field of view, or, in the case of spectroscopy, high spectral resolution across a widely dispersed spectrum. Missions to tackle this science with similar capabilities have been presented for several classes, from Flagship (Lynx) to Probe (AXIS) to Explorer (STAR-X) imaging missions, and high-resolution grating instruments flying aboard Flagship (Lynx XGS) and Probe/Explorer (Arcus) missions. These capabilities will be enabled by emerging technologies for lightweight, high-resolution X-ray optics and gratings\cite{Zhangetal2019,Heilmannetal2019}. 

\subsection{The need for a curved focal plane}

\begin{figure}[t]
\begin{center}
\includegraphics[width=.8\linewidth]{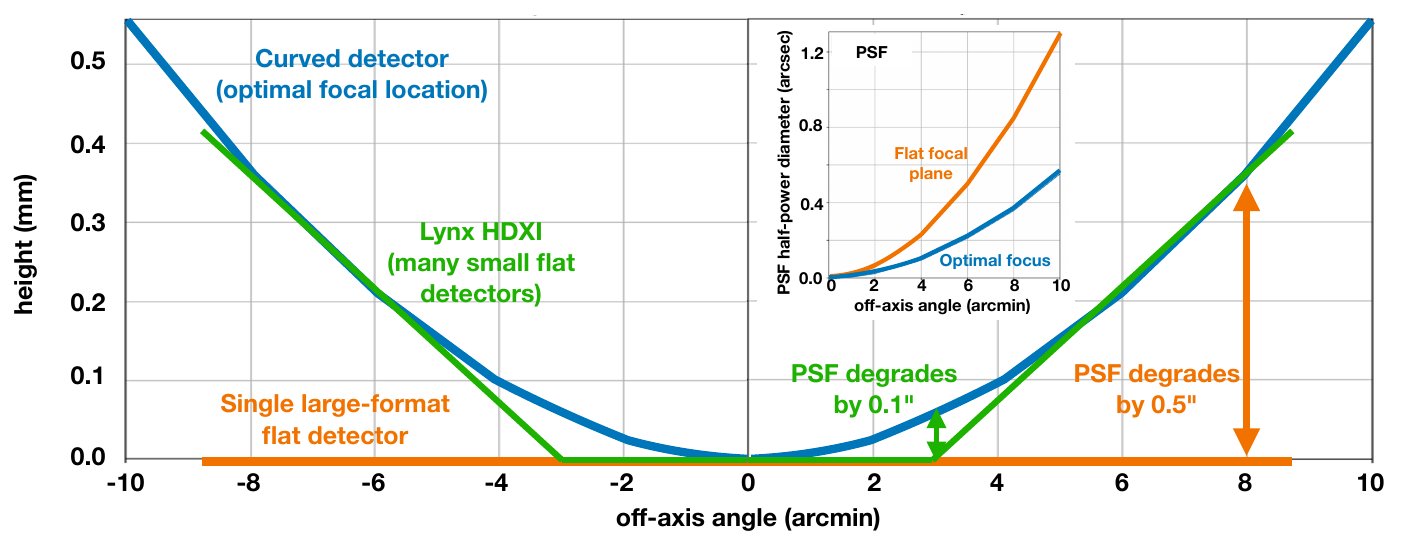}
\end{center}
\caption{Schematic of an advanced X-ray imager focal plane, placing three different detector layouts at the on-axis optimal focus. The inset compares the PSF of a flat focal plane to the optimal curved focal surface. Only a curved detector recovers the full resolving power of the optics at all off-axis angles.}
\label{fig:surface}
\end{figure}

An inherent feature of these advanced optical systems is that their surfaces of best focus are curved, not flat, with curvature radii as small as 1 to 2 m.  To fully exploit the capabilities of these sophisticated and expensive optics, and to maximize the science results outlined above, the associated instruments must provide detectors that closely conform to their curved focal surfaces. The power of an X-ray telescope is predicated on many factors, the most important of which are its point-spread function (PSF) and field of view (FoV). While on-axis PSF is determined by fabrication errors, off-axis PSF is in general determined by the mathematical prescriptions that define the basic design of an X-ray telescope. The grazing incidence nature of astronomical X-ray optics dictates that X-ray mirrors are nearly cylindrical in geometry, which means the optimal focal surface is curved. The effects of focal surface on image quality are shown in Figure \ref{fig:surface}, which results from detailed geometrical ray-tracing of a typical mirror shell of the Lynx mirror assembly.

The implications on future mission capabilities are profound. Flying a completely flat focal plane in an instrument like the High-Definition X-ray Imager (HDXI)\cite{HDXI} on Lynx would double the PSF size over half of the FoV, seriously compromising point-source sensitivity and survey power. Lynx has adapted for this by tilting the array of HDXI detectors to approximate the optimum focal surface, a technique used in previous instruments such as Chandra ACIS\cite{Garmireetal2003}. However, Lynx’s need for much larger area of sub-arcsecond imaging, with similar focal length and much smaller depth of focus, requires a large number of detectors, each of which is considerably smaller than scientific detectors that can be routinely fabricated today. This approach also requires detectors with four-side abuttability, complicates instrument design, and introduces undesirable gaps in precious high-resolution field coverage. Use of a curved detector greatly reduces this technical complexity and maximizes the return of the expensive, advanced optics.

\subsection{Curved detectors: the current and future landscape}

Curved detector technology has already been demonstrated in the ground-based DARPA/US Air Force Space Surveillance Telescope (SST), a system with spherically curved detectors of 5.44-m radius of curvature operating in the visible spectrum. These SST detectors, developed by MIT Lincoln Laboratory, have only recently been approved for public disclosure by the DoD.  The curved sensor technology used for SST was made possible by development of thinned, back-illuminated charge-coupled devices (CCDs)\cite{Westhoff2009,Burke2007}. We and others have since demonstrated that an imager can be fabricated in the planar state and then deformed to a curved surface\cite{Gregory2015,Gaschet2019}, and curved CCD and CMOS sensors have also appeared commercially\cite{Itonaga2014,Guenter2017,Joaquina2022}. This technology has the potential to improve field-averaged image quality, reduce the number of imaging detectors required to populate a focal surface, simplify detector and instrument design, and reduce the fraction of the field lost to inter-detector gaps. 

We here describe a joint effort between the MIT Kavli Institute for Astrophysics and Space Research (MKI) and MIT Lincoln Laboratory (MIT/LL) to exploit the demonstrated SST curved-CCD technology to provide large-area, high-performance detectors with the smaller radii of curvature required for future X-ray missions. The ultimate goal of this project is to to demonstrate that high-performance X-ray imaging detectors can be curved to match a focal surface similar to that envisioned for Lynx. Our specific objectives are to:
\begin{itemize}
    \item Curve an existing back-illuminated (BI) X-ray CCD (the MIT/LL CCID94), with diagonal length of 65 mm and active (fully depleted) thickness of 100 $\mu$m, to a radius of curvature of 2.5 m with an accuracy of 5 $\mu$m;
    \item Characterize detector properties (dark current, cosmetics, charge-transfer efficiency, noise, responsivity, X-ray spectral resolution, and quantum efficiency) of both flat and curved CCID94s to identify performance changes resulting from the deformation process; and  
    \item Expose curved and flat CCID94 detectors to notional (GEVS) vibration levels and representative thermal and radiation environments.  
\end{itemize}	

Our experience with SST has shown that there are challenges in fabricating curved sensors, including mechanical considerations such as deforming the brittle silicon or other semiconductor detector material without exceeding the fracture strain, introducing plastic deformation in metal layers that could lead to irreproducible characteristics, or buckling the deformed membrane into a shape that does not conform to the desired focal surface. It has been demonstrated that the deformation of the silicon can raise the dark current of the device, constituting a further challenge\cite{Gaschet2019}.

In this paper, we present details of the curving process and first results curving silicon samples that possess all the mechanical properties of the functional imagers that will soon be processed. We also present X-ray performance measurements of planar CCDs that will server as a baseline comparison to future testing of their curved counterparts.

\section{IMPLEMENTATION}

\subsection{Notional focal plane design and choice of test detector}

As an example of how a curved CCD array could be implemented, we adopt notional requirements of the Lynx Concept Study’s Design Reference mission\cite{Lynx}. The HDXI array would be about 65-mm across with focal-surface radius of curvature $\sim$2.5 m. As shown in Figure \ref{fig:surface}, using a simple flat imaging surface (orange) results in large departures from the optimal focus over the majority of the FoV. The reference design of the HDXI accounts for this by using 21 tiled imagers in a square 5$\times$5 array (with the outermost four corner chips missing)\cite{HDXI}. While significantly better, this approach requires four-side abuttable detectors, introduces a large number of chip gaps, and suffers additional seam loss due to square tiles not abutting tightly on a sphere; this effect increases with successive row of detectors. In addition, there remains a residual $\sim$0.1\arcsec\ PSF degradation across large parts of the FoV, an important contribution to the PSF budget for a sub-arcsecond wide-field imager.

A curved array for Lynx, on the other hand, could be fashioned from a 2$\times$2 array of three-side abutting CCDs, each with an imaging area about 33-mm square and with any circuitry to get the signal off-chip outside the field of view. It would also allow for the use of a framestore into which the exposed image could be transferred quickly and then read out more slowly, with less noise, and without a large contribution from out-of-time events. The seam loss of the array would be limited to a narrow stripe between the abutted CCDs, less than 0.5 mm across, producing less than 2\% loss.

\begin{figure}[t]
\begin{center}
\includegraphics[height=2.5in]{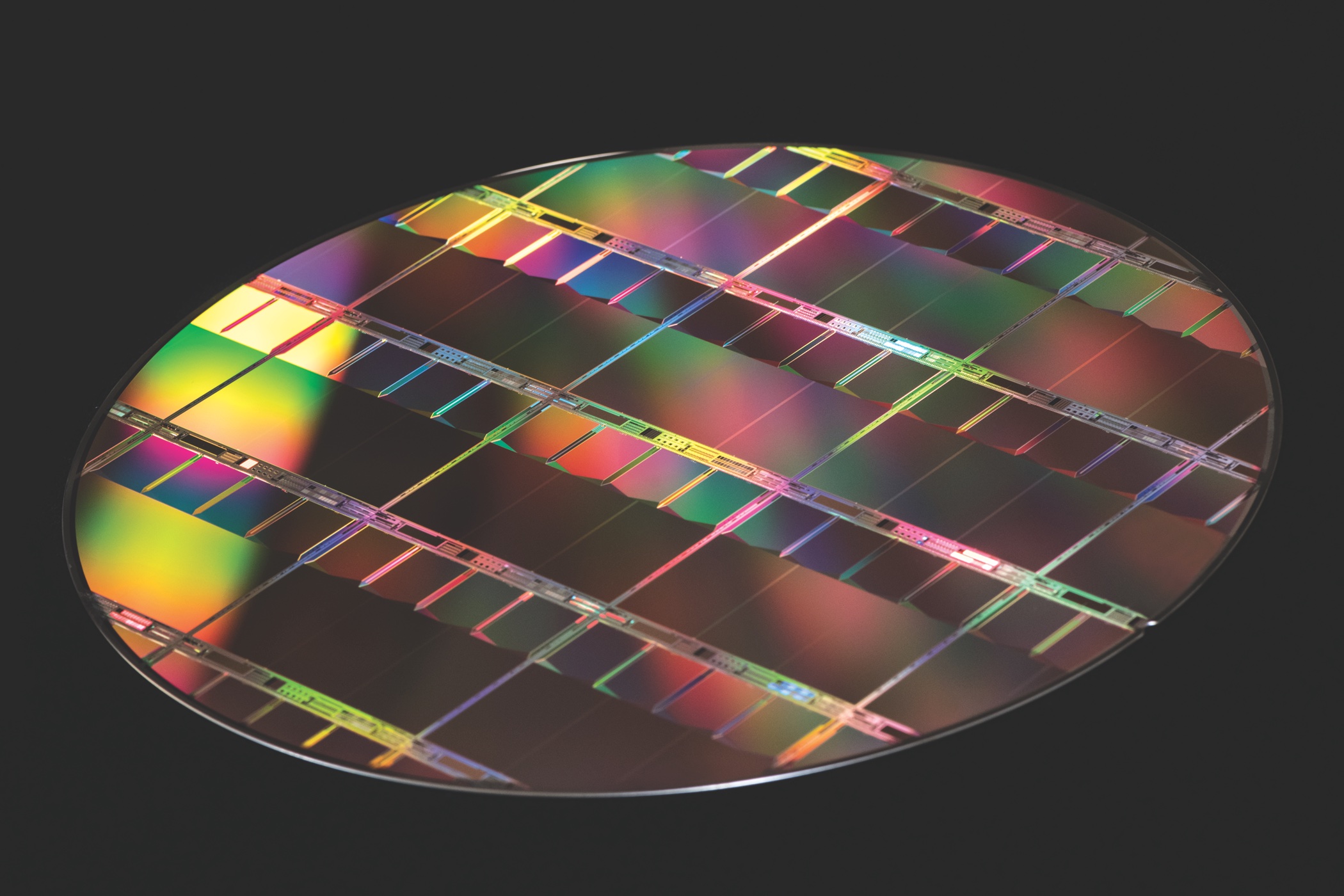}
\includegraphics[height=2.5in]{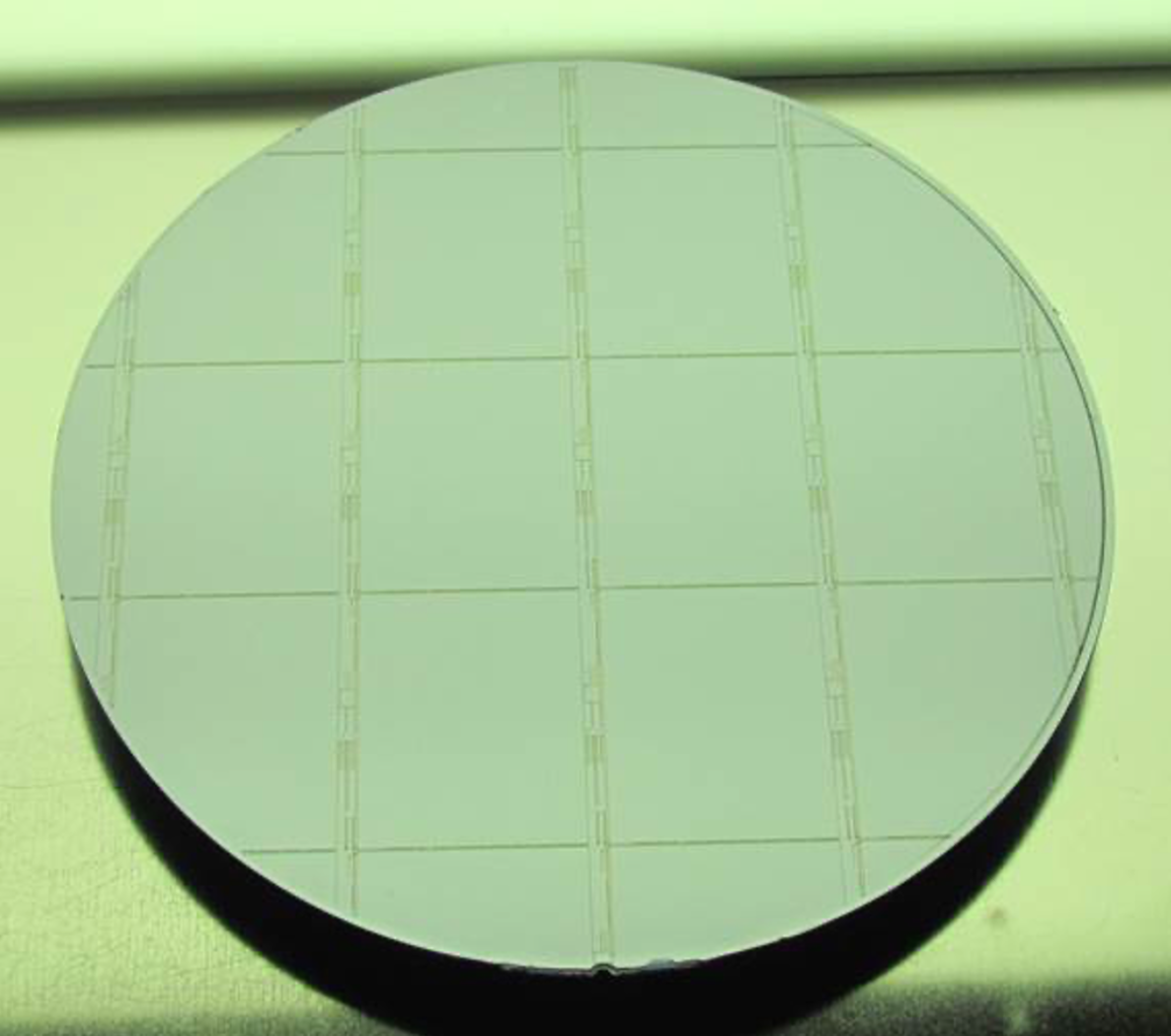}
\end{center}
\caption{Photographs of 200-mm wafers containing CCID94 CCDs through front-illumination (left) and back-illumination (right). Eight complete CCDs can be seen, each with a 50$\times$25-mm rectangular imaging area and four distinct framestore regions, the latter of which can only be seen in the front-illuminated devices. Each framestore feeds two outputs.}
\label{fig:ccid94_photo}
\end{figure}

The CCID94 detector developed by MIT/LL nearly fits this bill; it provides 1024 rows and 2048 columns of 24-$\mu$m pixels in the imaging area, or 24.6$\times$49.1 mm. The entire die including the adjacent framestore (with equal pixel count but smaller pixels) is 50.7 mm $\times$ 40.5 mm, or $\sim$65 mm along the diagonal. This diagonal extent is nearly identical to that for a square 33$\times$33-mm imaging-area chip, and so it provides a useful prototype on which to test the curving procedure. The CCID94, summarized in Table \ref{tab:ccid94}, has undergone years of development and fabrication at MIT/LL and extensive testing at MKI, and it is a well-understood, high-performance imager. It is baselined for the focal plane of the high-resolution X-ray grating spectrometer on the Arcus Explorer and Probe class mission concepts\cite{Smith2020_Arcus,Smith2023_Arcus,Smith2024_Arcus,Grant2024_Arcus}. Photographs of 200-mm wafers of CCID94 devices are shown in Figure \ref{fig:ccid94_photo}.

\begin{table}[t]
\caption{Features of the MIT/LL CCID94 CCD}\label{tab:ccid94}
\begin{center}       
\small
%%\begin{tabular}{|l|l|l|l|} %% this creates two columns
%% |l|l| to left justify each column entry
%% |c|c| to center each column entry
%% use of \rule[]{}{} below opens up each row
\begin{tabular}{|l|c|}
\hline\hline
\textbf{Feature}          & \textbf{CCID94} \\ \hline
Format                    & Frame-transfer, 2048$\times$1024 pixel imaging array \\
Image area pixel size     & 24$\times$24 $\mu$m \\
Output ports              & 8 pJFET                   \\
Transfer gate design      & Triple layer polysilicon  \\
Additional features       & Trough, charge injection  \\
BI detector thickness     & 50--100 $\mu$m            \\
Back surface              & MIT/LL MBE 5--10 nm          \\
Typical serial rate       & 0.5 MHz                   \\
Typical parallel rate     & 0.1 MHz                   \\
Full-frame read time      & $<$1 s                       \\
\hline\hline
\end{tabular}
\end{center}
\end{table} 

\subsection{Back illumination, surface passivation, curving, and packaging}

The approach taken to deform the wafers includes what is referred to as the double-flush thinning (DFT) process, so as to have sufficient flexibility for mounting on a curved silicon mandrel. In principle, this process is a continuation of BI methods routinely performed and described in detail previously\cite{Westhoff2009}, but presented here briefly.  A finished FI CCD wafer is inverted on a carrier wafer and the single-crystal silicon backside of the CCD etched in an acid solution to a thickness between 15 and 100 $\mu$m, depending on the needs of the application. The resulting backside is passivated as described below. Due to the very high resistivity of the CCD silicon and the imposed clock biases, the electric field extends through layers up to and exceeding 100 µm. This processing yields a nearly 100\% collection efficiency of photoelectrons. After backside passivation, several lithography and etch steps are needed to access the metal pads buried under oxide to enable device testing and packaging. Additional masking steps may also be used for light shields and/or anti-reflective coatings.

A critical aspect of detecting and resolving low-energy X-rays is being able to collect photoelectrons very near the surface (e.g., for 200-eV X-rays, the absorption distance is about 60 nm) and drive them towards the buried channel of the detector. This requires very good passivation of the incident surface, so it does not act as an electrical ``dead layer'', generating or recombining electrons. Consequently, a high QE requires a nearly perfect illuminated surface on the silicon, with a large electric field present to drive the photoelectrons away from that surface. The best-known way is to introduce a highly doped p-Si layer (Si-B) on the surface using molecular-beam epitaxy (MBE)\cite{Ryuetal2018}. MBE has an advantage over other methods of epitaxial growth of a single crystal in the steep profile of the high boron concentration, which creates the electric field necessary to drive the electrons away from the illuminated surface. 

\begin{figure}[t]
\begin{center}
\includegraphics[width=.5\linewidth]{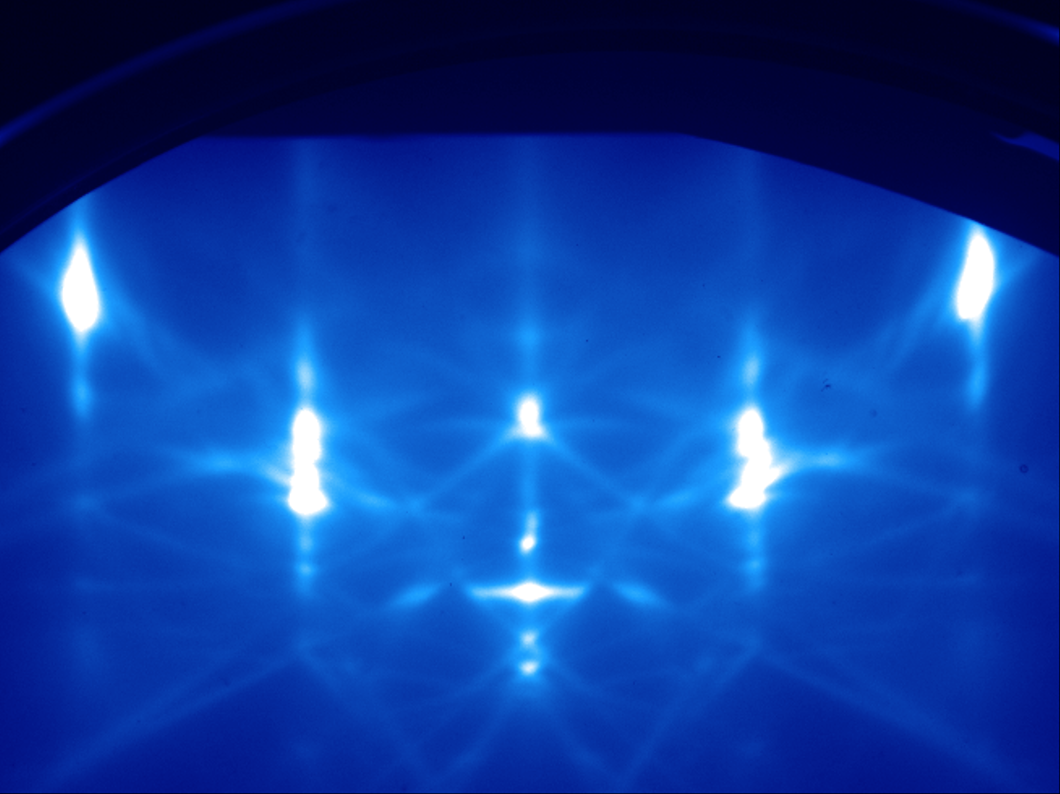}
\end{center}
\caption{Reflection high-energy electron diffraction (RHEED) pattern of an epitaxial MBE layer, exhibiting high crystallinity.}
\label{fig:RHEED}
\end{figure}

The MBE process consists of inserting a thinned CCD wafer in the MBE chamber, with the illuminated side facing the source of silicon and boron for the epitaxial growth. In this project, the wafers are thinned to 100 $\mu$m, in order to have good absorption of X-rays up to $\sim$10 keV. Cleaning of the wafer prior to entering the MBE system is critical\cite{Ryuetal2018}. The boron-containing layer is only a few nm thick, to ensure minimal photon absorption in the MBE entrance window. The quality of the epitaxial layer is determined in-situ using reflection high-energy electron diffraction (RHEED). An example of the diffraction pattern is shown in Figure \ref{fig:RHEED}. The bright points are due to particular lattice planes and their sharpness and the good definition of the arcs connecting them indicate a high-quality lattice.  As we describe below, the curving-process fixturing makes intimate contact with the detector back-surface during the curving process. Moreover, to our knowledge, no MBE-treated device has yet been curved. Therefore, a key objective of our project is to demonstrate that the detector can be curved without compromising the excellent X-ray performance of an MBE-treated device. 

\begin{figure}[t]
\begin{center}
\includegraphics[width=.85\linewidth]{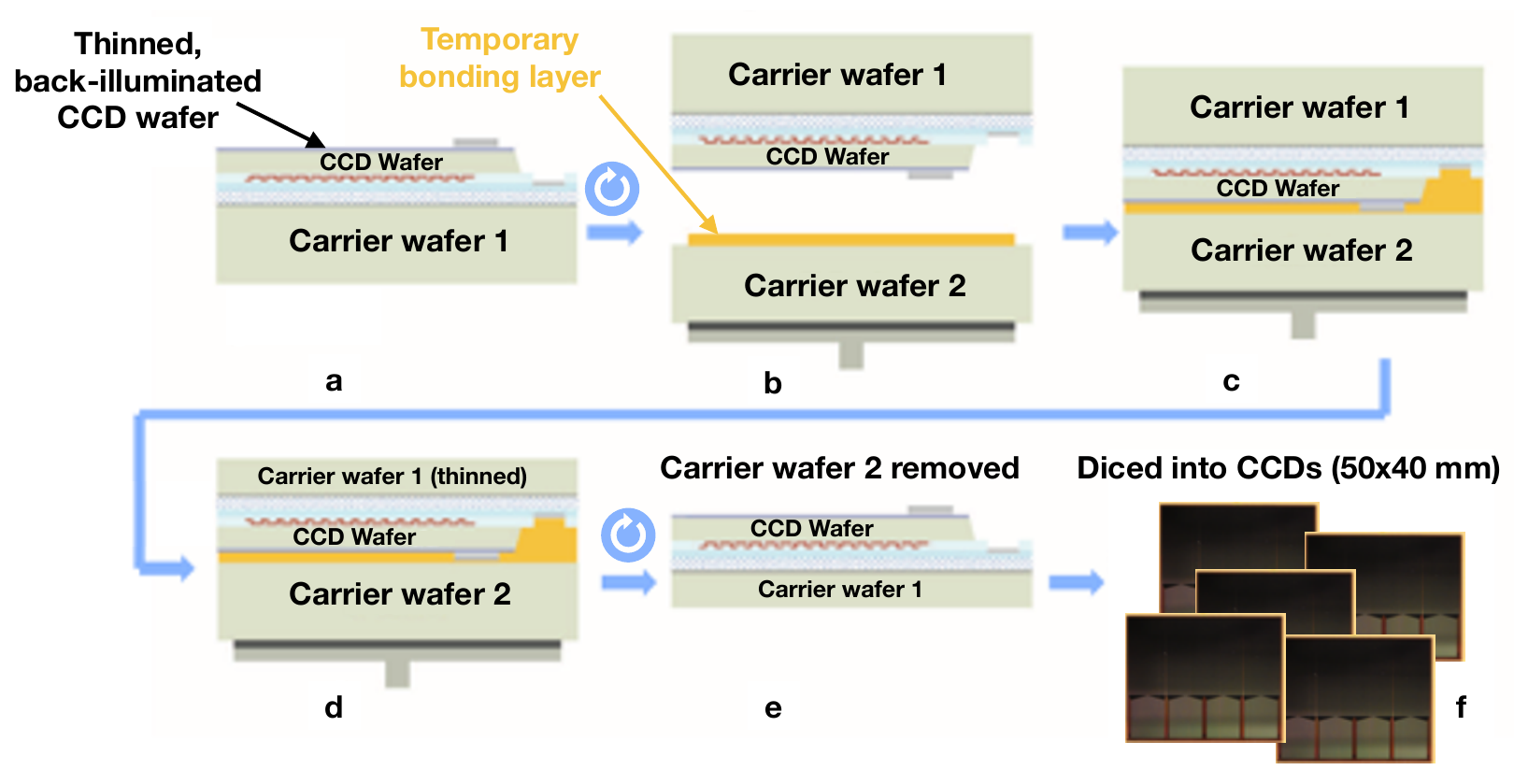}
\end{center}
\caption{The Double Flush Thinning (DFT) process to prepare a back-illuminated wafer for curving. a) Starting BI CCD wafer on a Si carrier wafer.  b) Inverting BI wafer and preparing second, temporary Si carrier wafer. c) Bonding BI and carrier wafers. d) Thinning of first carrier wafer. e) Removing temporary carrier wafer. f) Dicing into CCD devices (FI CCDs are shown here for clarity).}
\label{fig:dft}
\end{figure}

Once the standard BI processing with MBE passivation is completed, the wafer is ready for the DFT process (Figure \ref{fig:dft}a), where the illuminated surface is mounted temporarily to a second carrier wafer (Figure \ref{fig:dft}b and c) and the first carrier wafer is thinned so the device/handle composite is about 200 $\mu$m thick (Figure \ref{fig:dft}d). The second carrier is released, leaving the 200-$\mu$m thick composite (Figure \ref{fig:dft}e), which allows the 200-mm diameter wafer to be diced (Figure \ref{fig:dft}f) for individual CCD curving and mounting. This structure offers a very good compromise between strength and flexibility, allowing the CCDs to be deformed to meter-class radii with over 95\% yields. 

\begin{figure}[t]
\begin{center}
\vspace*{\baselineskip}
\includegraphics[width=.90\linewidth]{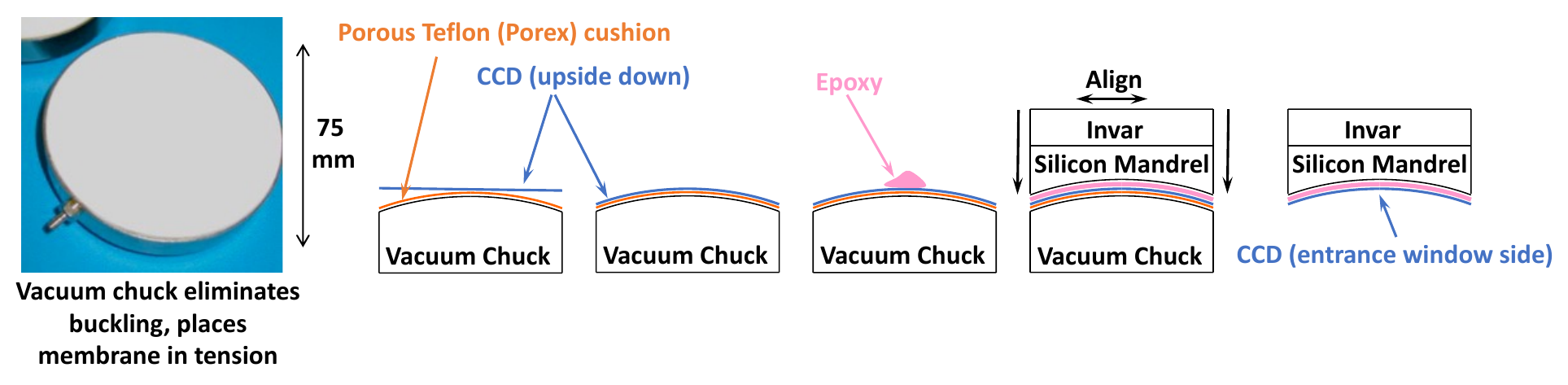}
\end{center}
\caption{Schematic of the curving process for a DFT-processed and diced CCD. The steps are described in the text.}
\label{fig:curving}
\end{figure}

The process to curve and permanently mount diced DFT chips to mandrels is based on the SST procedure, with two key differences: the X-ray CCD curvature is concave whereas the SST CCDs are convex; and radius of curvature for the X-ray CCDs is 2.5 m, whereas for SST it is 5.44 m. The vacuum-chuck bonding technique is largely the same, however, and this is illustrated in Figure \ref{fig:curving}. First a protective porous Teflon foil (shown as an orange layer) is placed on the convex, milled portion of the ceramic vaccum chuck to protect the illuminated surface of the CCD. The DFT CCD is placed illuminated-side down over the chuck and a vacuum applied, drawing the CCD into the appropriate curvature. Epoxy is placed on the now-convex rear surface of the thinned handle silicon and a  concave silicon mandrel is brought into contact and aligned using a specialized jig. The mandrels for SST were fabricated commercially to a spherical figure with surface accuracies of $\sim$1 $\mu$m, and we expect the same for the X-ray CCD mandrels. The mandrel and silicon are brought together with a large but non-destructive force, using a hydraulic plunger. Based on the SST experience, we expect to maintain this vacuum for several days to ensure proper curing of the epoxy between the imager and the mandrel. The concave CCD, now permanently mounted on its mandrel, is released from the chuck, wirebonded to a flexprint, and subjected to final testing. 

On SST, there have been no observed performance consequences due to curving the chips. Mechanical yield of the curving was around 95\%, and the induced strain was too low to detect a measurable increase in dark current. Even for radii of 1 m, the strains are quite low. To a first approximation, the maximum strain is $t/[2R(1-\nu)]$, where $t$ is the thickness of the composite body, $R$ is the radius of curvature, and $\nu$ is Poisson’s ratio. In cases where the sag approaches or exceeds $t$, an additional term $3D^2/64R^2$ must be added\cite{Prescott1946}, where $D$ is the diagonal of the deformed body. For this X-ray imager project, those terms sum to 0.003\%, well within the elastic limit for silicon.

\subsection{Current progress: fabricating and curving samples}

As a first step in the process, several different silicon ``dummy'' samples have been prepared for trial deformation. Two sample wafers, listed in Table \ref{tab:samples} as W\#, were prepared in a similar way to wafers intended for imaging devices, including components thought to be most prone to mechanical deformation effects; these include the metal conduction layers around the periphery of the CCD and the interlayer dielectric oxide within the active region. The dummies avoid the doping implants required for use as an imager, and also lack the triple polysilicon layers and associated oxide dielectric used for transfer gate electrodes. While the polysilicon will have a lower fracture stress than the single crystal silicon due to grain boundaries and other defects, it should be less fragile than the oxide layers that are included. This practical approach tests most mechanical components in the samples, but defers testing of all components for the functional imagers.

\begin{table}[t]
\caption{Description of wafer samples processed for deformation testing.}\label{tab:samples}
\begin{center}       
\scriptsize
% the >{\raggedright\arraybackslash} in front of the p{} (paragraph) column type forces left justification for lines that are wrapped
\begin{tabular}{|p{0.5in}|p{1.5in}|>{\raggedright\arraybackslash}p{2.0in}|>{\raggedright\arraybackslash}p{2.0in}|}
\hline\hline
94bi\_lot0 wafer ID &
 FI Lot &
 Substrate \& FI Process &
 BI Process \\ \hline
W1 &
 94BE\_TEST\_190064 Wafer 1 &
 P-Type (100) silicon monitor with two layers of metallization and attendant inter-layer dielectric &
 oxide-bonded, 100$\mu$m flush-thinned, DFT-processed with 150$\mu$m support \\ \hline
W2 &
 94BE\_TEST\_190064 Wafer 2 &
 P-Type (100) silicon monitor with two layers of metallization and attendant inter-layer dielectric &
 oxide-bonded, 100$\mu$m flush-thinned, DFT-processed with 150$\mu$m support \\ \hline
P2 &
 P-Type (100) silicon Wafer &
 P-Type (100) silicon Monitor &
 150$\mu$m flush-thinned (no support wafer) \\ \hline
P3 &
 P-Type (100) silicon Wafer &
 P-Type (100) silicon Monitor &
 250$\mu$m flush-thinned (no support wafer) \\ 
\hline\hline
\end{tabular}
\end{center}
\end{table} 

The W1 and W2 wafers underwent back-illumination processing in the MIT/LL Microelectronics Laboratory (ML)\footnote{\url{https://www.ll.mit.edu/about/facilities/microelectronics-laboratory}} as part of the CCID94 back-illumination lot that will also produce imagers for this project. The wafers have undergone the entire DFT process shown in Figure \ref{fig:dft}, including oxide bonding to a carrier wafer which was then thinned to 150 $\mu$m to produce the expected total thickness of the real detectors. 
%W9 and W12 were only flush-thinned to 100 $\mu$m, with no carrier wafer mounting.  
To allow additional testing with lower investment, two bare silicon wafers (P2 and P3) were thinned in the ML to thicknesses between 100 and 250 $\mu$m. All wafers in Table \ref{tab:samples} were diced into sample devices for curving.

\begin{figure}[t]
\begin{center}
\includegraphics[width=.3\linewidth]{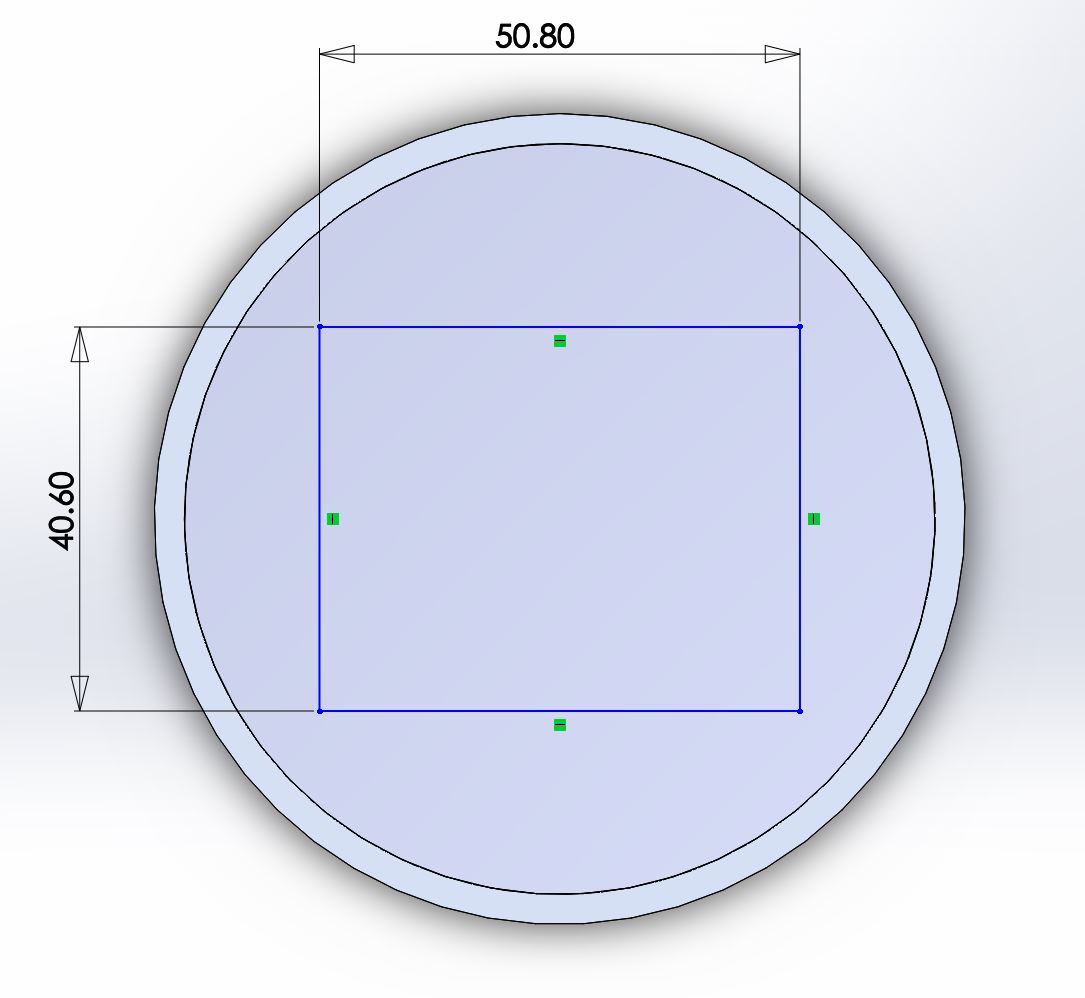}
\includegraphics[width=.4\linewidth]{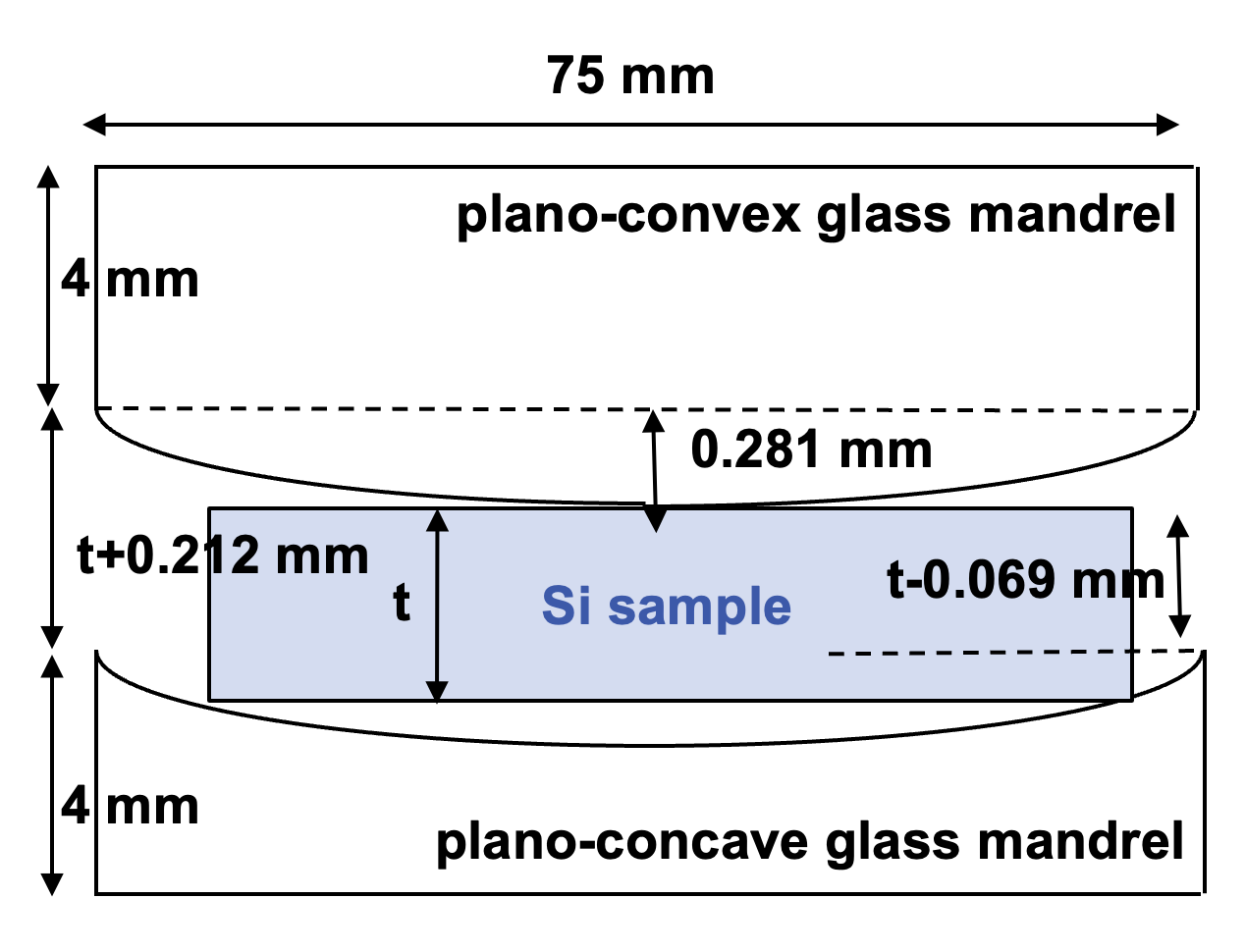}
\end{center}
\caption{(left) Top view of a rendering of the ceramic vacuum chuck. The inset rectangle shows the size of the CCD die that is placed on the chuck for curving. (right) Schematic showing two glass mandrels, one convex and one concave, in position to curve a silicon sample as anticipated for future testing. The dimensions are not shown to scale, and the shims referred to in the text are not shown. }
\label{fig:mandrels}
\end{figure}

Sample deformation testing was performed using a ceramic vacuum chuck that was designed and fabricated for this project. It has a porous convex surface with radius of curvature $\sim$2200 mm and diameter of 75 mm (see the photograph in Figure \ref{fig:curving}). As shown in Figure \ref{fig:mandrels}, this size is sufficient to hold the 50$\times$40-mm silicon samples and CCDs. The radius of curvature deviates from the intended 2500-mm value for reasons that are under invesigation. Samples were tested by simply placing them on the chuck and then applying a modest vacuum of 24 inHg, so that the ambient atmosphere imposes about 80 kPa of pressure; this conforms the sample to the chuck. The samples were inspected visually for buckling, then loaded into a Cyber Technologies CT300 non-contact profilometer and scanned with a 633-nm laser at 1-mm steps to verify spherical conformity of the curved sample. The chuck itself was also scanned with the CT300 to verify conformity.

To simulate the use of a silicon mounting mandrel in the curving process, two glass mandrel designs were fabricated, one plano-convex and one plano-concave, with a spherical radius of 2500 mm and diameter of 75 mm. These match the anticipated curvature of the final detectors, and are sufficiently large to hold an entire silicon sample. A schematic of the mandrels is shown in Figure \ref{fig:mandrels}. The mandrels are made of window-grade SCHOTT N-BK7, chosen for its economy and ease of use. For these tests, the optical quality and thermal properties of the glass are immaterial as it is used at room temperature and as a mechanical mold rather than a precise optical element. The concave mandrel was used in some tests to ``cap'' the samples deformed to the chuck and profile with th CT300. As with the chuck, both mandrels were also profiled individually with the CT 300.

While only the vacuum chuck (capped and uncapped with the concave mandrels) was used in the tests accomplished so far, in the future we will test-curve the samples between mandrels. The procedure for this is as follows (see Figure \ref{fig:mandrels}). First, an aluminum ring is placed around the plano-concave (lower) mandrel to secure it. The silicon sample of thickness $t$ is placed in the center of the mandrel, and shims are placed on top of the mandrel at 90\arcdeg\ intervals to protect the sample during this setup stage. These shims have thickness greater than $(t+0.212)$ mm to account for the mandrel curvature and inset of the sample, ensuring that no part of the plano-convex (upper) mandrel contacts the sample while the upper mandrel is slowly lowered and aligned onto the shims. The shims are then smoothly removed around the circumference, allowing the upper mandrel to contact the sample and deform it onto the lower mandrel. Once the shims are removed, the sample is inspected and photographed through the glass mandrels from both sides to inspect for buckling, and loaded into the CT300 for profilometry. 

For the P2 150-$\mu$m bare silicon sample, a layer of porous Teflon (Porex) was placed between the sample and the vacuum chuck. In the curving of real detectors, the interface will involve the fragile MBE-treated entrance window, which can be easily damaged by the ceramic. The Porex layer is intended to protect the MBE as well as provide an elastic membrane to produce better overall conformity. We expect to include this layer in future testing of samples.  We also note that while epoxy will be used to permanently mount the real detectors to a dedicated silicon mandrel, all of the testing described here is dry, without the use of epoxy.

\subsection{Test results}

\begin{figure}[p]
\begin{center}
\includegraphics[width=.8\linewidth]{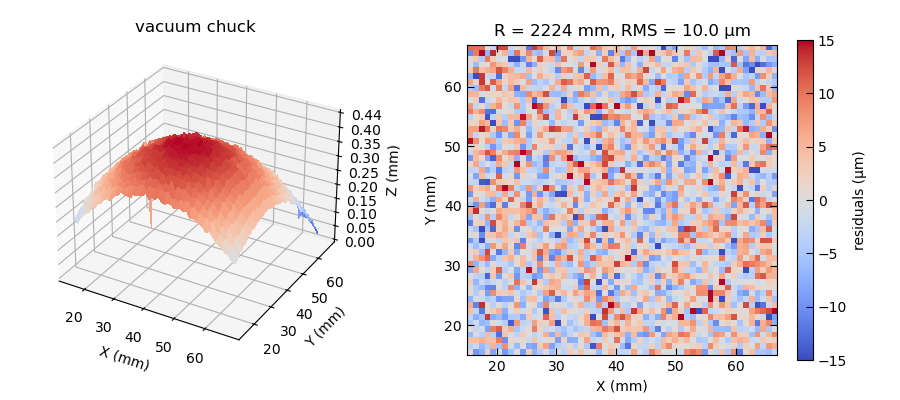}
\includegraphics[width=.8\linewidth]{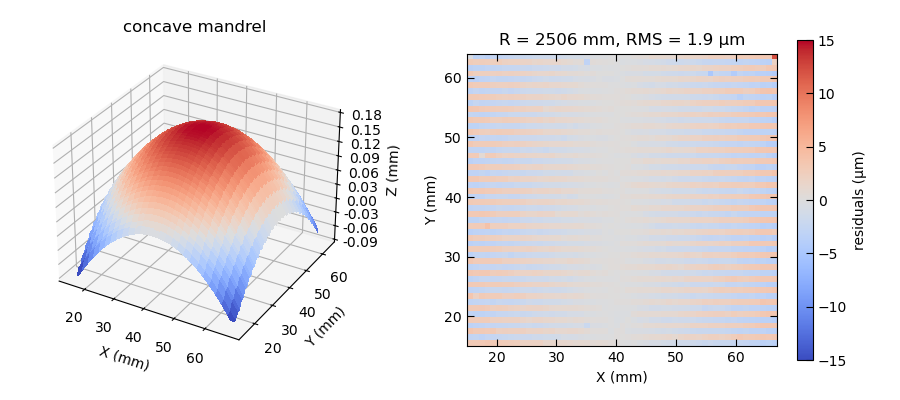}
\includegraphics[width=.8\linewidth]{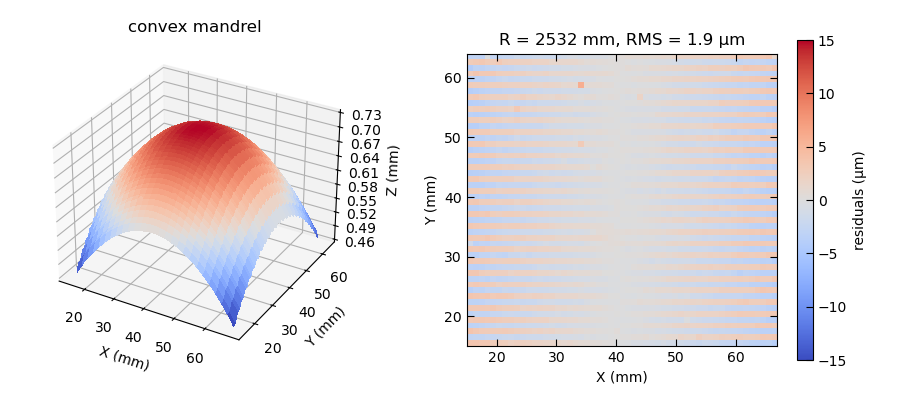}
\end{center}
\caption{(left panels) Surface profiles of the curved sides of the ceramic vacuum chuck and glass mandrels. Concave and convex surfaces are plotted in the same orientation for ease of visual 3-D representation. (right panels) Residuals after fitting a sphere to each surface profile. The spatial binning is equivalent to the scanner step size, 1 mm. The vacuum chuck has significant small-scale roughness on the order of 10 $\mu$m, while the glass mandrels are much smoother. There are no apparent large-scale departures from sphericity in all cases. Horizontal stripes are due to the boustrophedonic nature of the scanning.} 
\label{fig:surffit_mandrels}
\end{figure}

The bare vacuum chuck and glass mandrels show excellent spherical conformance, as shown in Figure \ref{fig:surffit_mandrels}. The glass mandrels in particular have RMS departures $<$2 $\mu$m from perfect smoothness, with a radius of curvature of 2500 mm, as expected. The ceramic vacuum chuck is considerably rougher, with RMS deviations of 10 $\mu$m on scales $\leq$1 mm, the resolution of the scanner. However, the radius of curvature of the chuck is considerably smaller than expected at 2200 mm. The cause of this smaller curvature is under investigation, but as a result the concave mandrel cannot be used to ``cap'' any samples on the chuck, as the osculating spheres would lead to buckling and other unexpected deformations. For this reason, we do not report results from capped samples in this work.

\begin{figure}[p]
\begin{center}
\includegraphics[width=.8\linewidth]{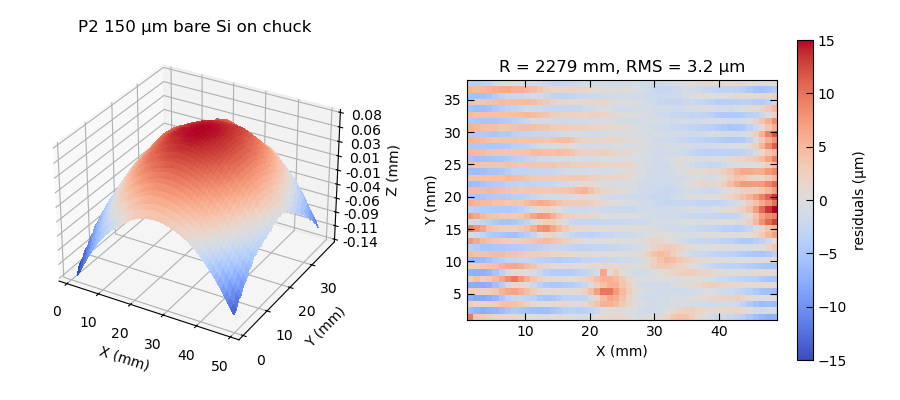}
\includegraphics[width=.8\linewidth]{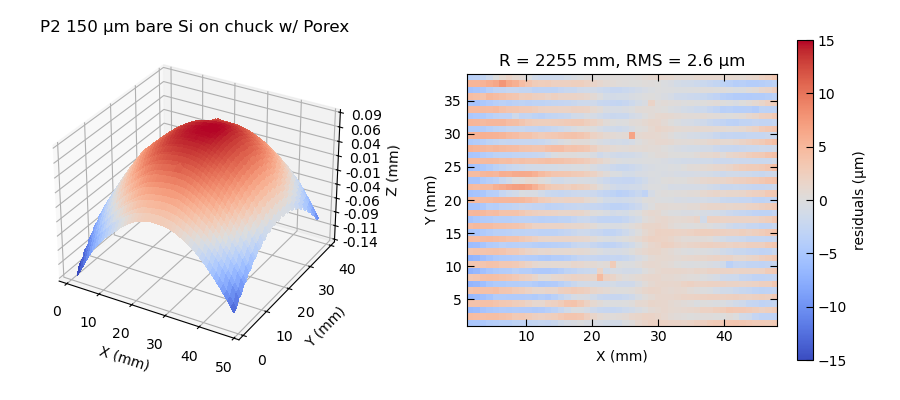}
\includegraphics[width=.8\linewidth]{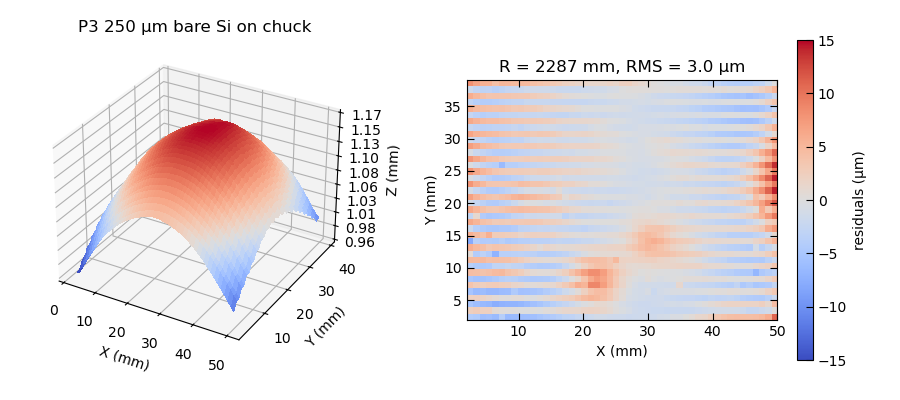}
\end{center}
\caption{Same as Figure \ref{fig:surffit_mandrels}, but for the P2 150-$\mu$m (top and middle) and P3 250-$\mu$m (bottom) thick bare silicon samples mounted on the vacuum chuck. The middle panel includes a Porex membrane between the sample and chuck, substantially decreasing the non-conformity at all scales.}
\label{fig:surffit_p13}
\end{figure} 

\begin{figure}[p]
\begin{center}
\includegraphics[width=.8\linewidth]{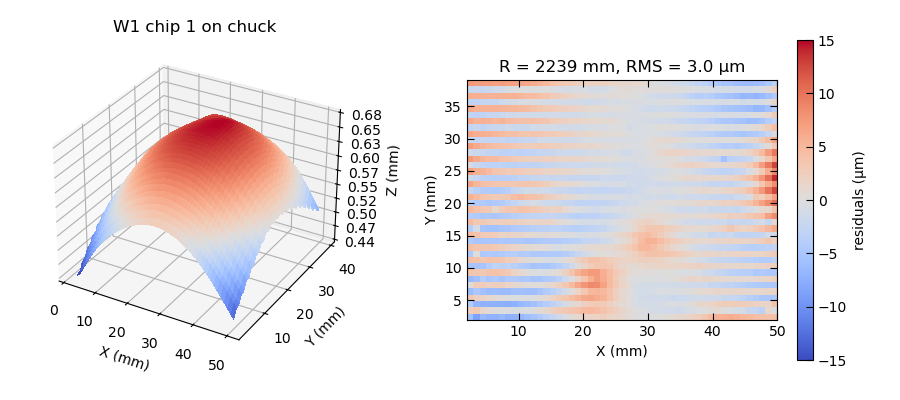}
\includegraphics[width=.8\linewidth]{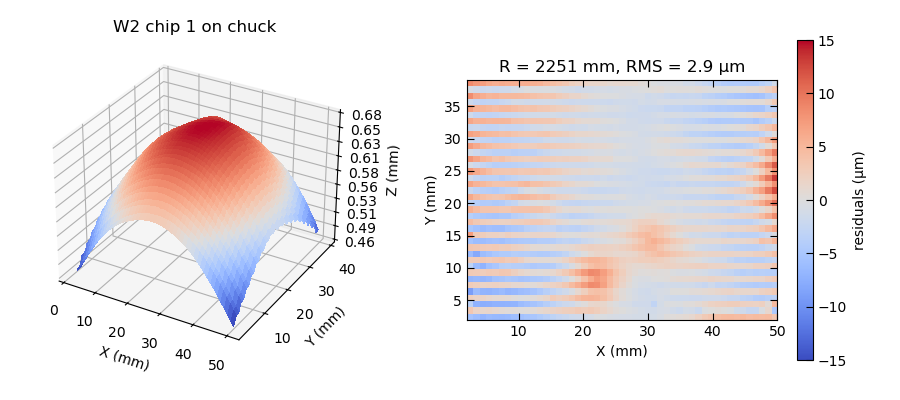}
\end{center}
\caption{Same as Figure \ref{fig:surffit_mandrels}, but for chips from the W1 (top) and W2 (bottom) sample wafers mounted on the vacuum chuck.}
\label{fig:surffit_w12}
\end{figure} 

Our initial testing produces good results at spherical curving of silicon samples on the vacuum chuck. Bare silicon samples of 150-$\mu$m and 250-$\mu$m thickness are deformed spherically to $\sim$3 $\mu$m RMS accuracy, as shown in Figure \ref{fig:surffit_p13}). Use of the Porex membrance between the sample and chuck improves the accuracy substantially, smoothing out large-scale ($\geq$5 mm) deviations that are common to all the measurements of un-cushioned samples. The two wafer chip samples produce similar results. Further improvements will include achieving a stronger vacuum, widespread use of the Porex layer as a cushion, and more thorough cleaning of the samples, chuck, and mandrels. Improvements will also be made in fabrication of the vacuum chuck to meet the specified radius of curvature.

\section{CCD performance testing}

The object of backside passivation is to meet the demanding requirements for the X-ray detection efficiency and spectral resolution imposed on future X-ray missions, particularly at X-ray energies below 1 keV.  For example, modern, high-resolution diffraction grating spectrometers, such as those recently proposed for the Arcus Probe mission\cite{Smith2023_Arcus} and developed as part of the Lynx mission concept study\cite{Lynx}, aim to detect and resolve lines of ionized carbon and oxygen (CVI, OVII and OVIII) at redshifts up to $z = 0.3$ or more. This requires high sensitivity to X-rays with wavelengths (energies) in the range 1.9 to 4.4 nm (650--280 eV). In addition, because these grating spectrometers achieve high spectral resolving power ($R = \lambda/\Delta\lambda \approx $ 3,000--10,000) by operating in high diffraction orders (as high as $m = 10$), the spectral resolution of the readout detector must be sufficient to resolve the energies of overlapping orders, which requires a detector resolving power of order $1/m$\cite{Gunther2023_Arcus_SPIE}. In the case of wide-field imaging instruments, the need to detect X-ray sources in the early Universe at redshifts as high as  $z \approx 10$ also imposes stringent demands on X-ray response at the extreme low-energy limit of the passband, 150--200 eV. Here, good detection efficiency requires not only good low-energy quantum efficiency, but also good spectral resolution to ensure that most of the charge deposited by the lowest-energy photons is recovered and distinguished from any noise peak. 

Understanding the effects of the curving process on CCD performance requires a detailed understanding of the baseline performance of uncurved CCDs. Toward this end, we have tested two 50-$\mu$m-thick back-illuminated CCID94 devices fabricated by MIT/LL as part of a different project. While not the same thickness as the CCDs that will be curved, the results provide a useful reference and guidance for future testing of 100-$\mu$m-thick planar and curved CCDs. We here present results from one device, given the designation CCID94L1W19C5 among the MKI group. Additional results have been previously presented\cite{Miller2023_AXIS} and are presented elsewhere in these proceedings\cite{LaMarr2024_SPIE}.

\begin{figure}[t]
\begin{center}
\includegraphics[width=.6\linewidth]{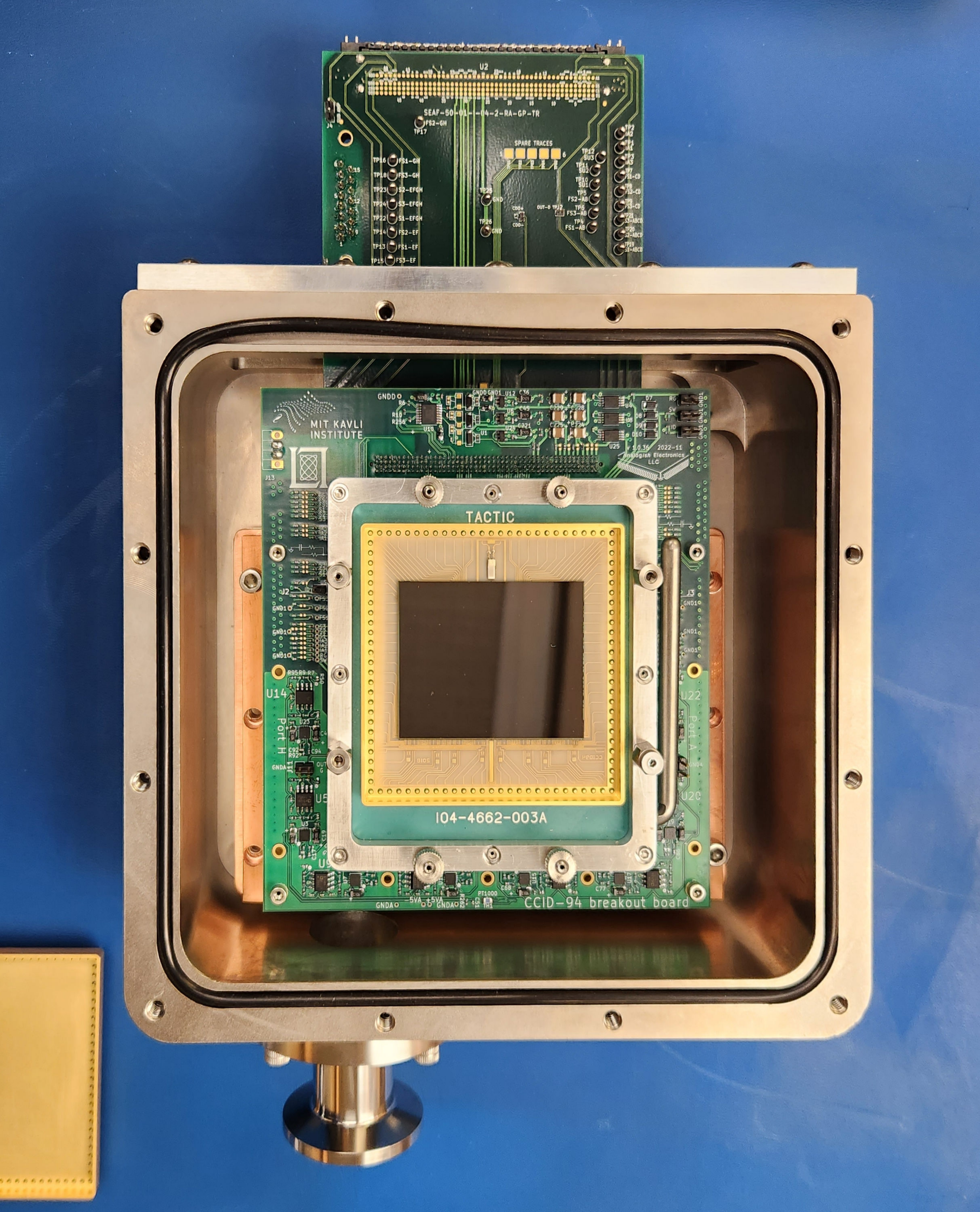}
\end{center}
\caption{Photograph of a packaged MIT/LL CCID94 back-illuminated CCD in its vacuum housing for testing at MKI.}
\label{fig:94labphoto}
\end{figure}

The test facilities at MKI include dedicated vacuum chambers for each type of CCD under testing, each equipped with a liquid-nitrogen cryostat for thermal control to temperatures below $-100$\arcdeg C. Each setup includes an Archon\footnote{\url{http://www.sta-inc.net/archon}} controller that provides CCD bias and clock voltages and performs digital sampling on the CCD analog video waveform, incorporating an interface board that connects to a custom made vacuum feed-through board and detector board for each CCD. The lab detector package and boards for each CCD were designed by MIT/LL. Each chamber allows easy insertion of a radioactive $^{55}$Fe source for reliable full-frame illumination with Mn K$\alpha$ (5.9 keV) and K$\beta$ (6.4 keV) X-rays. The CCID94 setup can incorporate a $^{210}$Po source with Teflon target that produces fluorescence lines of C K (0.27 keV) and F K (0.68 keV) across the full 50$\times$25 mm imaging area. For some testing presented here, the CCID94 chamber was mounted on an In-Focus Monochromator (IFM)\cite{Hettrick1990_ifm} that uses grazing incidence reflection gratings to produce clean monochromatic lines at energies below 2 keV, with typical spectral resolving power $\lambda/\Delta\lambda = E/\Delta E \sim $ 60--80, far higher than that of the CCD itself. We also mounted the chamber in the MKI polarimetry lab beamline\cite{Marshall2024_SPIE} for testing in a different environment, providing additional measurements at C K and O K (0.53 keV). A photo of the packaged CCD mounted in its test chamber is shown in Figure \ref{fig:94labphoto}.

All test data is acquired as a set of full pixel frames that are processed in a similar way to flight data on Chandra. Briefly, each image is corrected for pixel-to-pixel and time-dependent bias levels. Persistent (``hot'') pixels are masked, and local maxima above a defined event threshold (5--8 times the RMS noise) are identified. Pixel islands around these maxima are searched for pixels above a second ``split'' threshold (3--4 times the RMS noise), and all such pixels are summed to produce the event pulse height, a measurement of the incident photon energy. This summed pulse height is denoted ``allaboveph'' in several figures in this work. Pixel islands are 3$\times$3 for the CCID94 devices with 24-$\mu$m pixels. Each event is assigned a ``multiplicity'', akin to the ``grade'' on ASCA, Chandra, and Suzaku, but here simply encoding the number of pixels ``n\#'' that are above the split threshold.

\begin{figure}[p]
\begin{center}
\includegraphics[width=.45\linewidth]{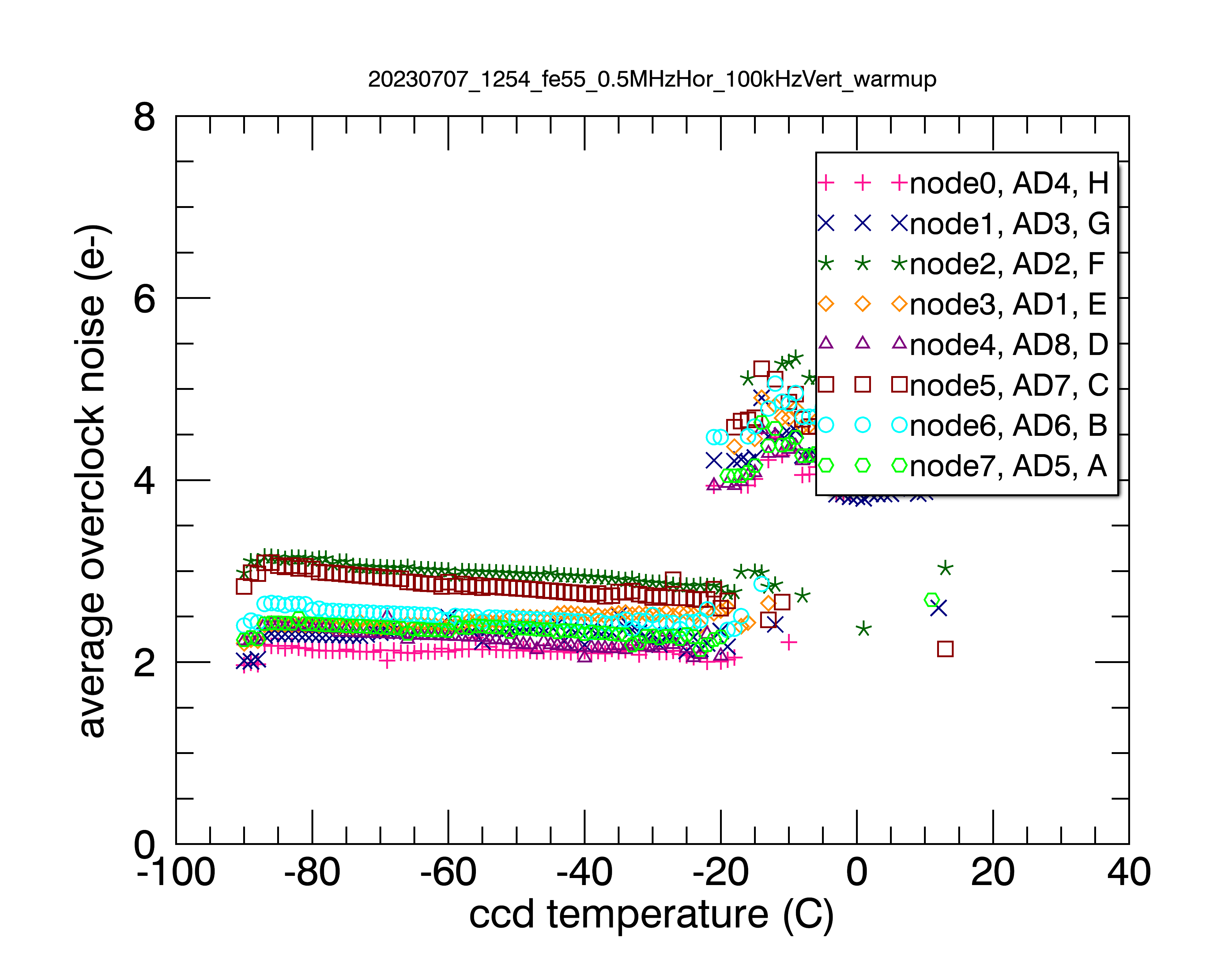}
\includegraphics[width=.45\linewidth]{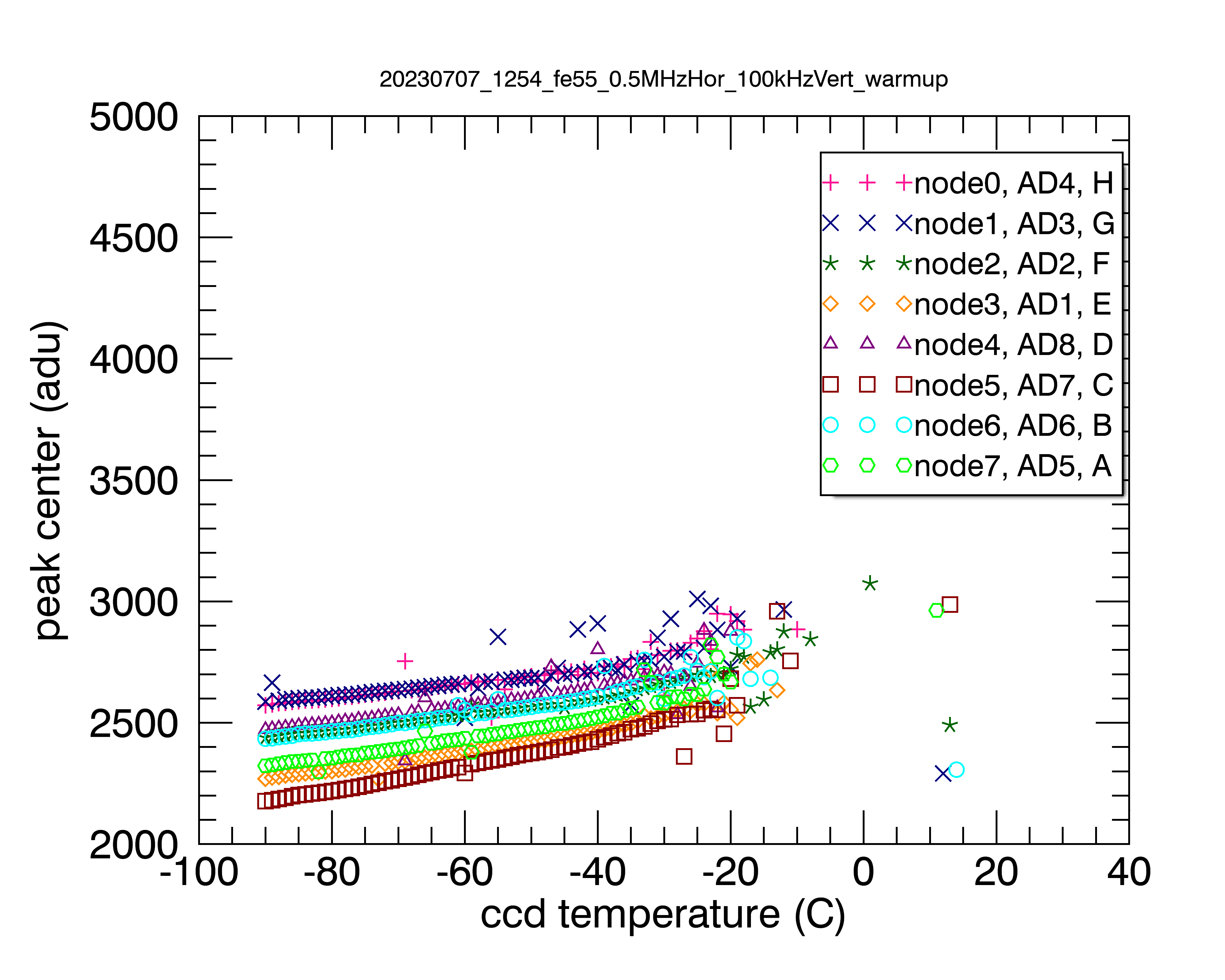}
\end{center}
\caption{(left) Readout noise as a function of temperature for all eight output nodes on the CCID94. (right) Peak in instrument units (ADU) of the Mn K$\alpha$ line at 5.9 keV. Both measurements are excellent and behave uniformly across segments and with temperature below $-20$\arcdeg C, where dark current ceases to dominate.}
\label{fig:94noise_gain}
\end{figure}

Basic performance of the planar CCID94 is excellent, as can be seen in Figure \ref{fig:94noise_gain}. The detector exhibits low noise that is uniform across the eight outputs and varies smoothly with temperature at a readout speed of 0.5 MHz, corresponding to a $\sim$0.5-s frame time. The gain is also uniform and well behaved with temperature below $-20$\arcdeg C, where dark current becomes low.

\begin{figure}[p]
\begin{center}
\includegraphics[width=.90\linewidth]{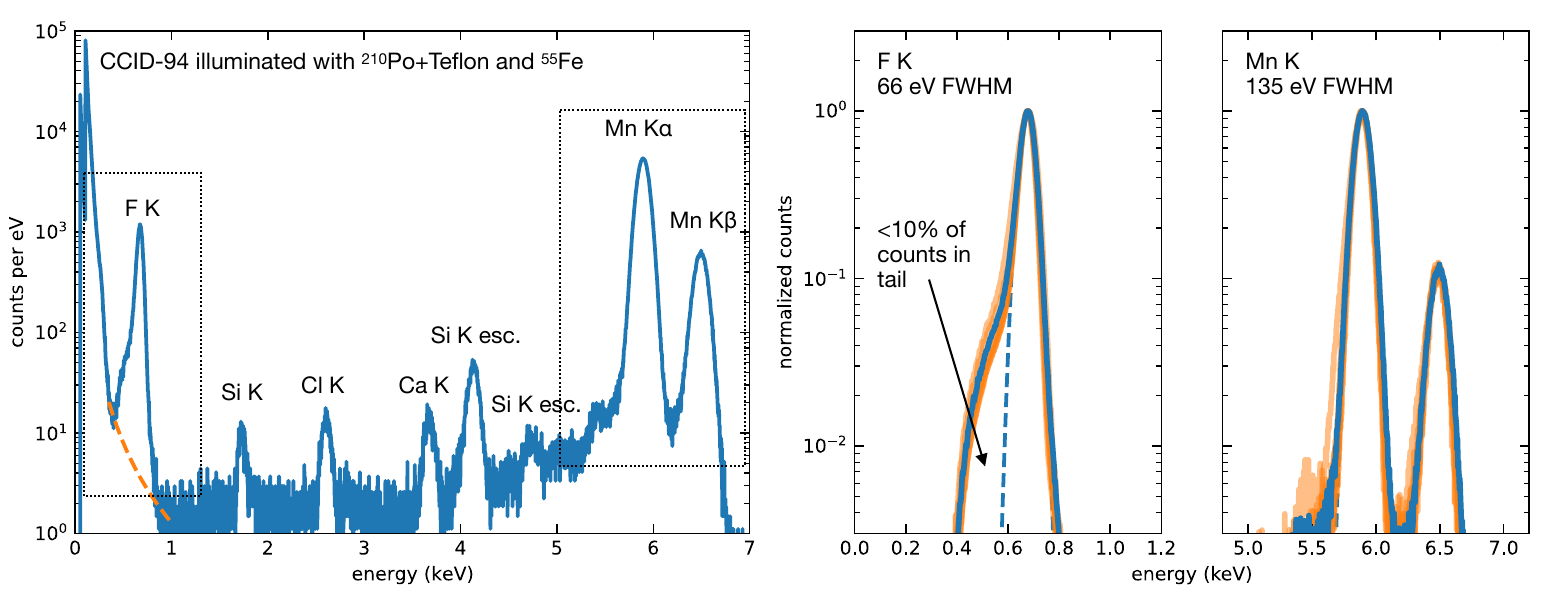}
\end{center}
\caption{(left panel) Spectrum of a single CCID94 segment (node `C') simultaneously illuminated with $^{210}$Po with a Teflon target and $^{55}$Fe. The primary fluorescence lines of F K, Mn K$\alpha$, and Mn K$\beta$ can be seen along with several other fluorescence and escape features. (right two panels) Zoom-in of the F K and Mn K peaks from the left panel, now also showing spectra from the other seven nodes of this chip as orange curves. The F K line has been corrected for the noise continuum by subtracting a best-fit power law, shown in dashed orange in the left panel. Gaussian fits (dashed blue lines) indicate excellent spectral FWHM at both energies. While the F K peak shows a non-Gaussian tail, it contains fewer than 10\% of the line counts.}
\label{fig:94pofe55}
\end{figure} 

We illuminated our test CCID94 device with $^{210}$Po with Teflon target and $^{55}$Fe simultaneously to produce a broad-band spectrum from a single representative segment (see Figure \ref{fig:94pofe55}).  The primary fluorescence lines of F K, Mn K$\alpha$, and Mn K$\beta$ can be seen along with several other fluorescence and escape features. C K is not visible here due to a low-energy noise peak that is unrelated to the detector or source. After fitting and subtracting a power-law model to this noise continuum, we measure the spectral resolution at F K (0.68 keV) to be 66 eV FWHM for all event multiplicities for this segment, with little variation from segment to segment. There is a non-Gaussian tail, which remains under investigation but could result from backside surface losses. This tail contains less than 10\% of the counts for all segments, and its segment-to-segment uniformity indicates it can be accounted for using a standard redistribution matrix. A similar tail is seen in our testing of another MIT/LL device, the CCID89, which has the same backside treatment, depletion depth, and pixel pitch as the CCID94.\cite{Bautz2024_SPIE}. The response at $\sim$6 keV is excellent for all nodes, averaging 135 eV FWHM for all event multiplicities. Results are summarized in Table \ref{tab:MKIresults}.

%%%%% table
\begin{table}[t]
\caption{Results from planar CCID94 testing at MKI.}\label{tab:MKIresults}
\begin{center}       
\scriptsize
\begin{tabular}{|ll|c|}
%% |l|l| to left justify each column entry
%% |c|c| to center each column entry
%% use of \rule[]{}{} below opens up each row
%\begin{tabular}{|p{1in}|>{\centering}p{1.2in}|>{\centering}p{1.2in}|>{\centering\arraybackslash}p{1.2in}|}
\hline\hline
\multicolumn{2}{|l|}{\textbf{Typical result}} & \textbf{CCID-94} \\ 
\hline
\multicolumn{2}{|l|}{Detector temperature}    &  $-89$\arcdeg    \\ 
\multicolumn{2}{|l|}{Serial readout rate}     &  0.5 MHz     \\ 
\multicolumn{2}{|l|}{Readout noise (RMS)}     &  2.1--3.1 e- \\ 
\multicolumn{2}{|l|}{Parallel CTI }        &  $< 10^{-6}$ per transfer\\ 
\hline
\multicolumn{3}{|l|}{Spectral resolution (Gaussian FWHM)$^a$} \\ 
\hline
C K           & all         &  68 (100\%) \\
0.27 keV      & n1          &  65 (51\%) \\
              & n2          &  70 (43\%) \\
              & n3          &  75 (5\%) \\
              & n4          &  60 (1\%) \\
\hline
O K           & all         & 57 (100\%)  \\
0.53 keV      & n1          & 53 (39\%) \\
              & n2          & 58 (48\%) \\
              & n3          & 69 (8\%) \\
              & n4          & 60 (5\%) \\
\hline
F K           & all         &  66 eV (100\%) \\
0.68 keV      & n1          &  63 eV (36\%)  \\
              & n2          &  66 eV (49\%)  \\
              & n3          &  80 eV (9\%)   \\
              & n4          &  74 eV (6\%)   \\
\hline
Mn K          & all          &139 eV (100\%) \\
5.9 keV       & n1           &129 eV (18\%)  \\
              & n2           &136 eV (44\%)  \\
              & n3           &140 eV (18\%)  \\
              & n4           &143 eV (20\%)  \\
\hline
\hline
\multicolumn{3}{l}{$^a$Spectral resolution is provided for each pixel multiplicity `n\#', with the } \\
\multicolumn{3}{l}{~fraction of events with that multiplicity given in parentheses.} \\
\multicolumn{3}{l}{~Multiplicities representing fewer than 1\% of events are not shown.} \\
\end{tabular}
\end{center}
\end{table} 

\begin{figure}[t]
\begin{center}
\includegraphics[width=.45\linewidth]{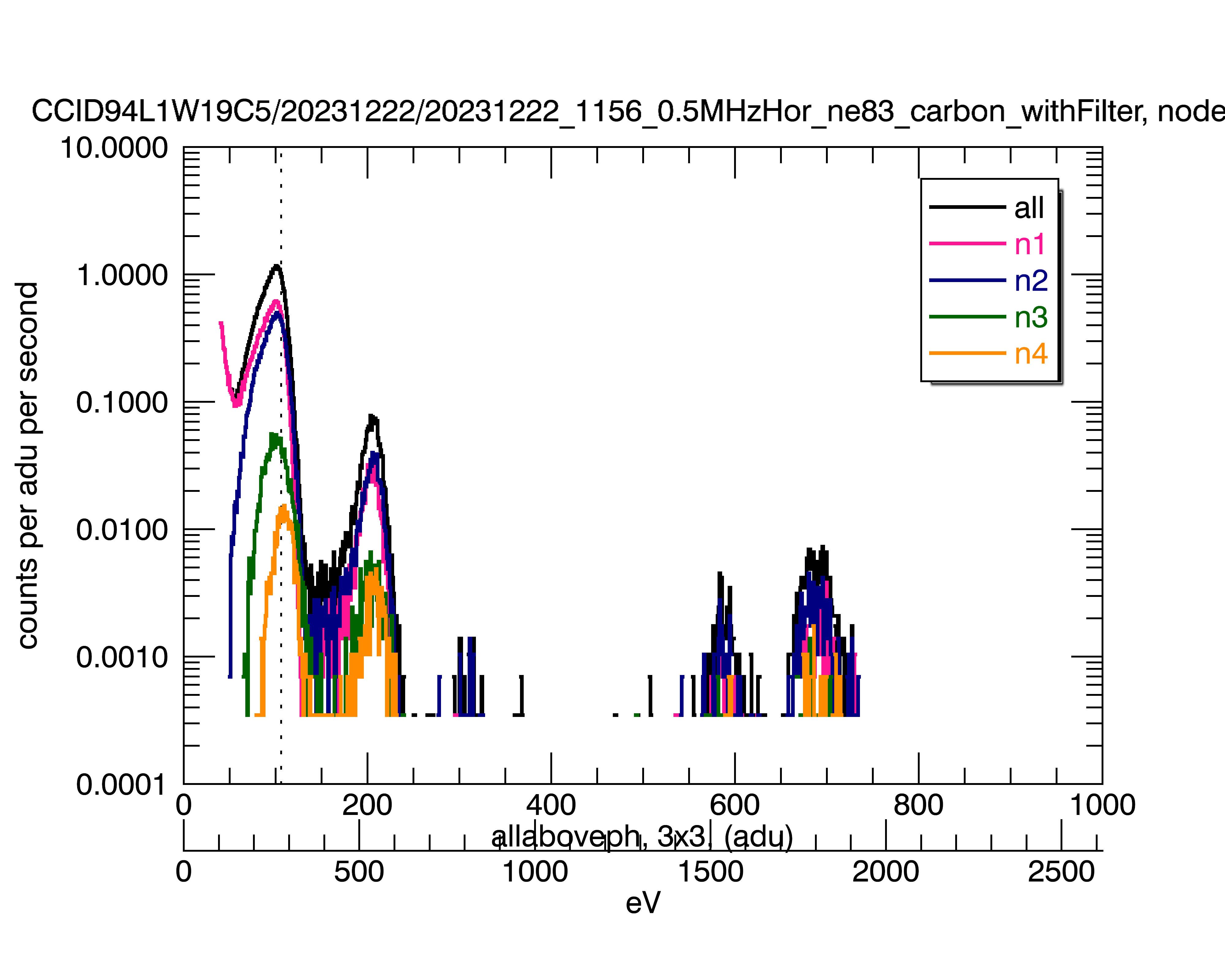}
\includegraphics[width=.45\linewidth]{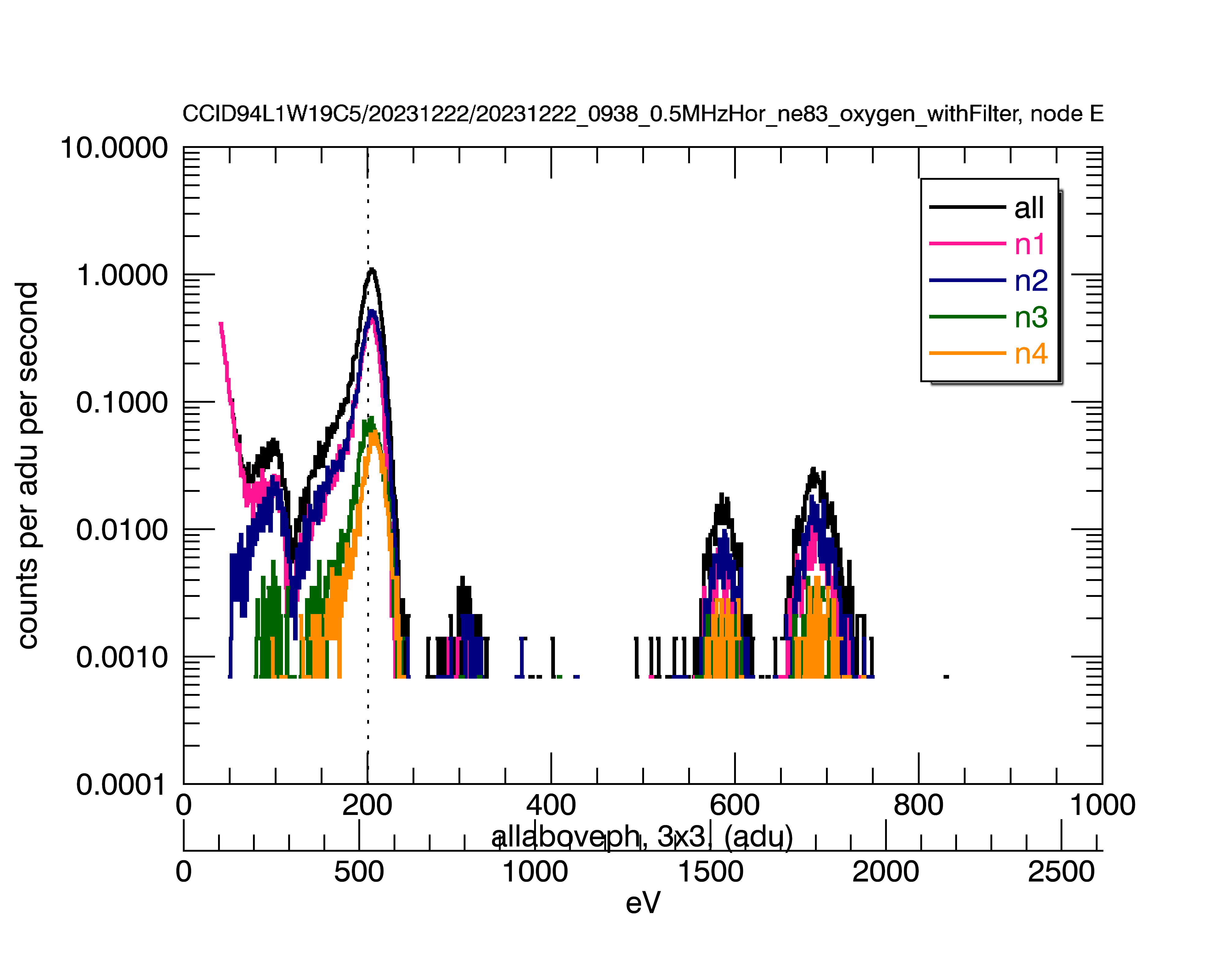}
\end{center}
\caption{Spectra of (left) C K and (right) O K obtained at the MKI polarimetry beamline for a representative segment on the CCID94. The dashed vertical lines show the expected line centers. Additional lines from different grating orders can be seen at low energies, along with fluorescence lines of Al K$\alpha$ (1.49 keV) and W M$\alpha$ (1.78 keV) produced by the polarizing multilayer holder and the multilayer itself, respectively.}
\label{fig:94pol}
\end{figure} 

Illumination with C K and O K fluorescence lines on the polarimetry beamline produced sufficient flux to overcome the low-energy noise peak, although only four of the eight segments were properly illuminated. We measured excellent response at 0.27 and 0.53 keV, as shown in Figure \ref{fig:94pol} and Table \ref{tab:MKIresults}.

\section{Summary}

Future X-ray imaging missions for astrophysics will require high spatial resolution over a wide field of view, however the cylindrical grazing-incidence optics employed produce a curved focal surface. To take full advantage of future advanced optics, we have begun a program to demonstrate that solid-state silicon imagers can be curved to match this surface, thereby reducing complexity with no effect on performance. Using bare silicon samples and wafers which have undergone metal layering, oxide bonding to a support wafer, and backside thinning, we show that our vacuum deformation technique is capable of creating a $\sim$2.5-m radius of curvature and meeting the 5-$\mu$m RMS conformance required. Use of a porous Teflon cushion improves the conformance substantially. We have also tested the baseline performance of planar CCDs to compare to the eventual curved functional devices, and report excellent X-ray response.

Future work will involve additional sample testing with improved cleanliness and correct figuring of the vacuum chuck as we move toward curving functional back-illuminated CCDs. We are also currently designing the package for the curved CCD, which requires substantial alteration from the flat detector package to incorporate the $\sim$7-mm-thick silicon mandrel. We anticipate completion of curving and testing our first functional CCDs within the next year.

\acknowledgments

We gratefully acknowledge support from NASA through Astrophysics Research and Analysis (APRA) grant 80NSSC22K0788.

% References
\bibliography{edm} % bibliography data in report.bib

\begin{thebibliography}{10}

\bibitem{Rees1984}
{Rees}, M.~J., ``{Black Hole Models for Active Galactic Nuclei},'' {\em
  \araa}~{\bf 22},  471--506 (Jan. 1984).

\bibitem{Volonteri2010}
{Volonteri}, M., ``{Formation of supermassive black holes},'' {\em \aapr}~{\bf
  18},  279--315 (July 2010).

\bibitem{Inayoshi2020}
{Inayoshi}, K., {Visbal}, E., and {Haiman}, Z., ``{The Assembly of the First
  Massive Black Holes},'' {\em \araa}~{\bf 58},  27--97 (Aug. 2020).

\bibitem{Rees1988}
{Rees}, M.~J., ``{Tidal disruption of stars by black holes of
  {}10$^{6}$-{}10$^{8}$ solar masses in nearby galaxies},'' {\em \nat}~{\bf
  333},  523--528 (June 1988).

\bibitem{Troja2018}
{Troja}, E., {Piro}, L., {Ryan}, G., {van Eerten}, H., {Ricci}, R., {Wieringa},
  M.~H., {Lotti}, S., {Sakamoto}, T., and {Cenko}, S.~B., ``{The outflow
  structure of GW170817 from late-time broad-band observations},'' {\em
  \mnras}~{\bf 478},  L18--L23 (July 2018).

\bibitem{Smith2011}
{Smith}, B.~D., {Hallman}, E.~J., {Shull}, J.~M., and {O'Shea}, B.~W., ``{The
  Nature of the Warm/Hot Intergalactic Medium. I. Numerical Methods,
  Convergence, and O VI Absorption},'' {\em \apj}~{\bf 731},  6 (Apr. 2011).

\bibitem{Bregman2018}
{Bregman}, J.~N., {Anderson}, M.~E., {Miller}, M.~J., {Hodges-Kluck}, E.,
  {Dai}, X., {Li}, J.-T., {Li}, Y., and {Qu}, Z., ``{The Extended Distribution
  of Baryons around Galaxies},'' {\em \apj}~{\bf 862},  3 (July 2018).

\bibitem{Smith2020_Arcus}
{Smith}, R.~K., ``{The Arcus soft x-ray grating spectrometer explorer},'' in
  [{\em Space Telescopes and Instrumentation 2020: Ultraviolet to Gamma
  Ray}{\nolinebreak\hspace{0.1em}]},  {den Herder}, J.-W.~A., {Nikzad}, S., and
  {Nakazawa}, K., eds., {\em Society of Photo-Optical Instrumentation Engineers
  (SPIE) Conference Series} {\bf 11444},  114442C (Dec. 2020).

\bibitem{Zhangetal2019}
{Zhang}, W.~W., {Allgood}, K.~D., {Biskach}, M.~P., {Chan}, K.-W., {Hlinka},
  M., {Kearney}, J.~D., {Mazzarella}, J.~R., {McClelland}, R.~S., {Numata}, A.,
  {Riveros}, R.~E., {Saha}, T.~T., and {Solly}, P.~M., ``{High-resolution,
  lightweight, and low-cost x-ray optics for the Lynx observatory},'' {\em
  Journal of Astronomical Telescopes, Instruments, and Systems}~{\bf 5},
  021012 (Apr. 2019).

\bibitem{Heilmannetal2019}
{Heilmann}, R.~K., {Bruccoleri}, A.~R., {Song}, J., and {Schattenburg}, M.~L.,
  ``{Progress in x-ray critical-angle transmission grating technology
  development},'' in [{\em Optics for EUV, X-Ray, and Gamma-Ray Astronomy
  IX}{\nolinebreak\hspace{0.1em}]},  {\em Society of Photo-Optical
  Instrumentation Engineers (SPIE) Conference Series} {\bf 11119},  1111913
  (Sept. 2019).

\bibitem{HDXI}
{Falcone}, A., {Kraft}, R., and {Bautz}, M., ``{The Lynx high defintion x-ray
  imager},'' in [{\em Space Telescopes and Instrumentation 2018: Ultraviolet to
  Gamma Ray}{\nolinebreak\hspace{0.1em}]},  {\em {Proc. SPIE}} {\bf 10699}
  (August 2018).

\bibitem{Garmireetal2003}
{Garmire}, G.~P., {Bautz}, M.~W., {Ford}, P.~G., {Nousek}, J.~A., and {Ricker},
  Jr., G.~R., ``{Advanced CCD imaging spectrometer (ACIS) instrument on the
  Chandra X-ray Observatory},'' in [{\em X-Ray and Gamma-Ray Telescopes and
  Instruments for Astronomy}{\nolinebreak\hspace{0.1em}]},  {J.~E.~Truemper \&
  H.~D.~Tananbaum}, ed., {\em \procspie} {\bf 4851},  28--44 (Mar. 2003).

\bibitem{Westhoff2009}
Westhoff, R.~C., Burke, B.~E., Clark, H.~R., Loomis, A.~H., Young, D.~J.,
  Gregory, J.~A., and Reich, R.~K., ``{Low dark current, back-illuminated
  charge coupled devices},'' in [{\em Sensors, Cameras, and Systems for
  Industrial/Scientific Applications X}{\nolinebreak\hspace{0.1em}]},  Bodegom,
  E. and Nguyen, V., eds.,  {\bf 7249},  72490J, International Society for
  Optics and Photonics, SPIE (2009).

\bibitem{Burke2007}
Burke, B.~E., Gregory, J.~A., Cooper, M.~J., Loomis, A.~H., Young, D.~J., Lind,
  T.~A., Doherty, P., Daniels, P., Landers, D.~J., Ciampi, J., Johnson, K.~F.,
  and O'Brien, P.~W., ``{CCD Imager Development for Astronomy},'' {\em Lincoln
  Laboratory Journal}~{\bf 16}(2),  392 (2007).

\bibitem{Gregory2015}
Gregory, J.~A., Smith, A.~M., Pearce, E.~C., Lambour, R.~L., Shah, R.~Y.,
  Clark, H.~R., Warner, K., III, R. M.~O., Woods, D.~F., DeCew, A.~E., Forman,
  S.~E., Mendenhall, L., DeFranzo, C.~M., Dolat, V.~S., and Loomis, A.~H.,
  ``Spherically curved ccd imagers for passive sensors,'' {\em OSA---The
  Optical Society}~{\bf 54}(10),  3072 (2015).

\bibitem{Gaschet2019}
Gaschet, C., Jahn, W., Chambion, B., Hugot, E., Behaghel, T., Lombardo, S.,
  Lemared, S., Ferrari, M., Caplet, S., G{\'e}tin, S., Vandeneynde, A., and
  Henry, D., ``Methodology to design optical systems with curved sensors,''
  {\em Applied Optics}~{\bf 58},  973 (01 2019).

\bibitem{Itonaga2014}
Itonaga, K., Arimura, T., Matsumoto, K., Kondo, G., Terahata, K., Makimoto, S.,
  Baba, M., Honda, Y., Bori, S., Kai, T., Kasahara, K., Nagano, M., Kimura, M.,
  Kinoshita, Y., Kishida, E., Baba, T., Baba, S., Nomura, Y., Tanabe, N.,
  Kimizuka, N., Matoba, Y., Takachi, T., Takagi, E., Haruta, T., Ikebe, N.,
  Matsuda, K., Niimi, T., Ezaki, T., and Hirayama, T., ``A novel curved cmos
  image sensor integrated with imaging system,'' in [{\em 2014 Symposium on
  VLSI Technology (VLSI-Technology): Digest of Technical
  Papers}{\nolinebreak\hspace{0.1em}]},   1--2 (2014).

\bibitem{Guenter2017}
Guenter, B., Joshi, N., Stoakley, R., Keefe, A., Geary, K., Freeman, R.,
  Hundley, J., Patterson, P., Hammon, D., Herrera, G., Sherman, E., Nowak, A.,
  Schubert, R., Brewer, P., Yang, L., Mott, R., and Mcknight, G., ``Highly
  curved image sensors: a practical approach for improved optical
  performance,'' {\em Optics Express}~{\bf 25},  13010 (05 2017).

\bibitem{Joaquina2022}
{Joaquina}, K., {Jahn}, W., {Struss}, Q., {Delcroix}, P., {Renard}, M.,
  {Cornu}, S., {Lemared}, S., and {Mehri}, T., ``{Curved CMOS imaging sensor:
  development and reliability test results},'' in [{\em Society of
  Photo-Optical Instrumentation Engineers (SPIE) Conference
  Series}{\nolinebreak\hspace{0.1em}]},  {Minoglou}, K., {Karafolas}, N., and
  {Cugny}, B., eds., {\em Society of Photo-Optical Instrumentation Engineers
  (SPIE) Conference Series} {\bf 12777},  127776S (July 2023).

\bibitem{Lynx}
{Gaskin}, J.~A., {Swartz}, D.~A., {Vikhlinin}, A., {{\"O}zel}, F., {Gelmis},
  K.~E., {Arenberg}, J.~W., {Bandler}, S.~R., {Bautz}, M.~W., {Civitani},
  M.~M., {Dominguez}, A., {Eckart}, M.~E., {Falcone}, A.~D.,
  {Figueroa-Feliciano}, E., {Freeman}, M.~D., {G{\"u}nther}, H.~M., {Havey},
  K.~A., {Heilmann}, R.~K., {Kilaru}, K., {Kraft}, R.~P., {McCarley}, K.~S.,
  {McEntaffer}, R.~L., {Pareschi}, G., {Purcell}, W., {Reid}, P.~B.,
  {Schattenburg}, M.~L., {Schwartz}, D.~A., {Schwartz}, E.~D., {Tananbaum},
  H.~D., {Tremblay}, G.~R., {Zhang}, W.~W., and {Zuhone}, J.~A., ``{Lynx X-Ray
  Observatory: an overview},'' {\em Journal of Astronomical Telescopes,
  Instruments, and Systems}~{\bf 5},  021001 (Apr. 2019).

\bibitem{Smith2023_Arcus}
{Smith}, R., ``{The Arcus probe mission},'' in [{\em UV, X-Ray, and Gamma-Ray
  Space Instrumentation for Astronomy XXIII}{\nolinebreak\hspace{0.1em}]},
  {Siegmund}, O.~H. and {Hoadley}, K., eds., {\em Society of Photo-Optical
  Instrumentation Engineers (SPIE) Conference Series} {\bf 12678},  126780E
  (Oct. 2023).

\bibitem{Smith2024_Arcus}
Smith, R.~K., ``{Arcus Probe: revealing feedback-driven structure and evolution
  throughout the universe},'' in [{\em Space Telescopes and Instrumentation
  2024: Ultraviolet to Gamma Ray}{\nolinebreak\hspace{0.1em}]},  den Herder,
  J.-W.~A., Nikzad, S., and Nakazawa, K., eds.,  {\bf 13093},  13093--81,
  International Society for Optics and Photonics, SPIE (2024).

\bibitem{Grant2024_Arcus}
Grant, C.~E., Bautz, M.~W., Miller, E.~D., Foster, R.~F., LaMarr, B.~J.,
  Malonis, A.~C., Prigozhin, G., Schneider, B., Leitz, C., and Falcone, A.~D.,
  ``{The Focal Plane of the Arcus Probe X-Ray Spectrograph},'' {\em Journal of
  Astronomical Telescopes, Instruments, and Systems}~{\bf submitted} (2024).

\bibitem{Ryuetal2018}
{Ryu}, K.~K., {Leitz}, C.~W., {Clark}, H.~R., {Chen}, X., {Cooper}, M.~J.,
  {Zhu}, M., {Welander}, P.~B., {Lambert}, R.~D., {Bolkhovsky}, V., {Yost}, D.
  R.~W., {Burke}, B.~E., {Gregory}, J.~A., and {Suntharalingam}, V.,
  ``{Oxide-bonded molecular-beam epitaxial backside passivation process for
  large-format CCDs},'' in [{\em Space Telescopes and Instrumentation 2018:
  Ultraviolet to Gamma Ray}{\nolinebreak\hspace{0.1em}]},  {den Herder},
  J.-W.~A., {Nikzad}, S., and {Nakazawa}, K., eds., {\em Society of
  Photo-Optical Instrumentation Engineers (SPIE) Conference Series} {\bf
  10699},  106993P (July 2018).

\bibitem{Prescott1946}
Prescott, J.,  [{\em Applied Elasticity}{\nolinebreak\hspace{0.1em}]}, Dover
  (1946).

\bibitem{Gunther2023_Arcus_SPIE}
G{\"u}nther, H.~M., Cheimets, P., Miller, E.~D., DeRoo, C., Smith, R.~K., Ptak,
  A., and Heilmann, R.~K., ``{Arcus x-ray telescope performance and
  alignment},'' in [{\em UV, X-Ray, and Gamma-Ray Space Instrumentation for
  Astronomy XXIII}{\nolinebreak\hspace{0.1em}]},  Siegmund, O.~H. and Hoadley,
  K., eds.,  {\bf 12678},  126781D, International Society for Optics and
  Photonics, SPIE (2023).

\bibitem{Miller2023_AXIS}
Miller, E.~D., Bautz, M.~W., Grant, C.~E., Foster, R., LaMarr, B., Malonis, A.,
  Prigozhin, G., Schneider, B., Leitz, C., Herrmann, S., Allen, S.~W.,
  Chattopadhyay, T., Orel, P., Morris, G.~R., Stueber, H., Falcone, A.~D.,
  Ptak, A., and Reynolds, C., ``{The high-speed x-ray camera on AXIS},'' in
  [{\em UV, X-Ray, and Gamma-Ray Space Instrumentation for Astronomy
  XXIII}{\nolinebreak\hspace{0.1em}]},  Siegmund, O.~H. and Hoadley, K., eds.,
  {\bf 12678},  1267816, International Society for Optics and Photonics, SPIE
  (2023).

\bibitem{LaMarr2024_SPIE}
LaMarr, B.~J., Schneider, B., Prigozhin, G., Miller, E.~D., Bautz, M.~W.,
  Foster, R.~F., Grant, C.~E., Malonis, A.~C., Cooper, M.~J., Lambert, R.~D.,
  Ryu, K.~K., and Jensen, M., ``{Soft X-ray resolution and scientific
  performance of CCD sensors for future X-ray missions},'' in [{\em X-Ray,
  Optical, and Infrared Detectors for Astronomy
  XI}{\nolinebreak\hspace{0.1em}]},  Holland, A.~D. and Minoglou, K., eds.,
  {\bf 13103},  13103--33, International Society for Optics and Photonics, SPIE
  (2024).

\bibitem{Hettrick1990_ifm}
Hettrick, M.~C., ``In-focus monochromator: theory and experiment of a new
  grazing incidence mounting,'' {\em Appl. Opt.}~{\bf 29},  4531--4535 (Nov
  1990).

\bibitem{Marshall2024_SPIE}
Marshall, H.~L., Heine, S., Garner, A., LaMarr, B.~J., and Schneider, B.,
  ``{Characterization of x-ray detectors in the MIT x-ray polarimetry
  beamline},'' in [{\em X-Ray, Optical, and Infrared Detectors for Astronomy
  XI}{\nolinebreak\hspace{0.1em}]},  Holland, A.~D. and Minoglou, K., eds.,
  {\bf 13103},  13103--64, International Society for Optics and Photonics, SPIE
  (2024).

\bibitem{Bautz2024_SPIE}
Bautz, M.~W., Miller, E.~D., Prigozhin, G., LaMarr, B.~J., Malonis, A., Foster,
  R.~F., Grant, C.~E., Schneider, B., Leitz, C., Donlon, K., Prigozhin, I.,
  Lambert, R.~D., Cooper, M.~J., Herrmann, S.~C., Orel, P., Chattopadhyay, T.,
  Morris, G.~R., Wilkins, D.~R., Stueber, H.~R., Poliszczuk, A., and Allen,
  S.~W., ``{Fast, low-noise X-ray image sensor technology for strategic X-ray
  astrophysics missions},'' in [{\em Space Telescopes and Instrumentation 2024:
  Ultraviolet to Gamma Ray}{\nolinebreak\hspace{0.1em}]},  den Herder,
  J.-W.~A., Nikzad, S., and Nakazawa, K., eds.,  {\bf 13093},  13093--63,
  International Society for Optics and Photonics, SPIE (2024).

\end{thebibliography}
\bibliographystyle{spiebib} % makes bibtex use spiebib.bst

\end{document}